\documentclass[aps,prd,twocolumn,superscriptaddress,floatfix]{revtex4-1}
\usepackage[pdftex]{graphicx}

\graphicspath{{figs/}}
\usepackage[makeroom]{cancel}
\usepackage{xcolor}
\usepackage{float}
\usepackage{amsmath,amssymb,amsfonts,amsbsy}
\usepackage{hyperref}  

\renewcommand{\Im}{\mathrm{Im }}

\begin{document}
	

\title{Instanton induced transverse single spin asymmetry for \texorpdfstring{$\pi^0$}{pi} production in \texorpdfstring{$pp$}{pp}-scattering}


\author{N.Korchagin}
 \email{korchagin@impcas.ac.cn}
 \affiliation{Institute of Modern Physics, Chinese Academy of Sciences, Lanzhou 730000, China}
 
\author{N.Kochelev}
\noaffiliation

\author{Pengming Zhang}
 \email{zhangpm5@mail.sysu.edu.cn}
  \affiliation{School of Physics and Astronomy, Sun Yat-sen University, Zhuhai 519082, China}


\begin{abstract}

We calculate the production cross-section and the transverse single-spin asymmetry for pion in $p^{\uparrow}+p\to \pi^0 + X$.
Our computation is based on existence of the instanton induced effective quark-gluon and quark-gluon-pion interactions with a strong spin dependency.
In this framework we calculate the cross section without using fragmentation functions.
We compare predictions of the model with data from RHIC.
Our numerical results, based on the instanton liquid model for QCD vacuum, are in  agreement with unpolarized cross section data. 
The asymmetry grows with the transverse momentum of pion $k_t$ in accordance with experimental observations. It reach value $\sim 10\%$ but at higher $k_t$ than experiment shows.

\end{abstract}


\maketitle


\section{Introduction}

Transverse single-spin asymmetries (TSSAs) have been puzzling physicists more than three decades.
They are among the most intriguing observables in hadronic physics since first FermiLab measurements for $p+\mathrm{Be} \to \Lambda^{\uparrow}+X$ reaction\cite{Bunce:1976yb}.
Since then, TSSA are observed in many different reaction, including mesons production in $pp$ and SIDIS.
Results of experiments are in contradiction with predictions from the perturbative Quantum Chromodynamics(pQCD) and the naive collinear parton model. It was expected that asymmetries should be extremely small\cite{Kane:1978nd}.
For comprehensive introduction to the problematic we refer the reader to review \cite{DAlesio:2007bjf}.
In this paper we focus on transverse single-spin asymmetry for pion production in nucleon–nucleon scattering. 
It is often called as analyzing power and denoted as $A_N$.
Such measurements were done at FermiLab by E581/E704 Collaborations\cite{Adams:1991cs}.
Later, similar measurements at higher energy was performed at RHIC\cite{Adams:2006uz}.
Unambiguous effects were measured and they triggered renewed interest on TSSAs.


A popular approach to describe observed spin effects is based on the extension of the collinear parton model with inclusion of parton's transverse motion.
It utilizes the Transverse Momentum Dependent(TMD) factorization scheme.
However, the factorization theorem has not been proven generally for such case\cite{Rogers:2010dm}.
It has so far only been proven for some classes of processes: the Drell-Yan 
$(q + \bar{q} \to l^{+} + l^{-})$\cite{Collins:1984kg} and semi-inclusive DIS\cite{Collins:1981va}.
$k_T$-dependent factorization is, therefore, an assumption, although a well-accepted one.
Efforts are ongoing to establish the theoretical basis more firmly.
We refer the reader to papers with discussions of the universality \cite{Collins:2002kn,Metz:2002iz,Bomhof:2004aw,Boer:2003cm} and TMD pdf's evolution\cite{Henneman:2001ev,Kundu:2001pk}.
Moreover, the dominance of $k_T$ effects among other contributions is disputed.
For example, effects of parton virtuality, target mass corrections could be of the same order of magnitude as transverse parton motion\cite{Moffat:2017sha}.


Two mechanisms for TSSA have been proposed in the framework of non-collinear parton model.
The first is Collins mechanism, when transversity distribution in combination with spin-dependent, chiral-odd Fragmentation Function(FF) can give rise TSSA\cite{Collins:1992kk}.
The Collins FF describes the azimuthal asymmetry of a fragmented hadron in respect to struck quark polarization.
Work \cite{Anselmino:2012rq,Ma:2004tr} has suggested that it is difficult to explain the large TSSA entirely in terms of the Collins effect.


The second mechanism was suggested by Sivers\cite{Sivers:1989cc}.
The idea is that parton distributions are asymmetric in the intrinsic transverse momentum $k_T$ within the proton.
The Sivers effect can exist both for quarks and gluons.
This intrinsic asymmetry is represented by Sivers function of the unpolarized partons in a transversely polarized proton.
Calculations based on Sivers effect for E704 data and other results can be found in \cite{Anselmino:2013rya,Anselmino:1994tv}.


Other direction for investigation is the twist-3 approach.
It was pointed out that three-parton correlators may give rise to TSSAs\cite{Efremov:1981sh}.
Qiu and Sterman examined higher-twist contributions due interference between quark and gluon fields in the initial polarized proton\cite{Qiu:1998ia}.
Similar study was performed by Kanazawa and Koike for quark-gluon interference in the final state\cite{Kanazawa:2000hz}.


In present paper we propose an alternative mechanism for TSSA in $pp\to\pi X$, based on existence of novel effective interaction induced by instantons.
The instantons describe sub-barrier transitions between the classical QCD vacua with different topological charges.
In previous work\cite{Kochelev:2013zoa} we calculated TSSA for quark-quark scattering and showed that such mechanism gives significant TSSA. 
However, generalization of that result to the case of real hadron scattering is unclear.
Calculation in the standard, pQCD-like way with introduction of fragmentation functions is not self-consistent. 
Extraction of FFs requires evolution equation and was done in framework of pQCD without considering an additional non-perturbative low-energy interaction.
The new vertex may give significant contribution to the evolution\cite{Kochelev:2015pqd}.
Reanalyzing data with the new vertex and modified evolution will not give new information since we will introduce more parameters.


Fortunately, the low-energy effective interaction generated by instantons provides us the other solution.
It contains a pion-quark-gluon vertex. 
In such case, we do not need any fragmentation function and, as result, we reduce the number of parameters in the model. 
Formation of pion happens at the short distance of the instanton scale $\approx 0.3$~fm, which is smaller than distances of confinement dynamics.
The other important consequence is breaking of the pQCD factorization. Scattering of partons and hadronization are coherent at the instanton scale. It might be a corner stone of various phenomena observed in high energy reactions in the few GeV range for the transferred momentum. 


This paper has the following structure.
Section \ref{section:vertex} gives a brief introduction to the instanton generated interaction.
In section \ref{section:Xsec} we discuss calculation for pion production cross-section and then, in section \ref{section:ssa}, calculate TSSA.
Section \ref{section:discussion} is dedicated to numerical analysis and discussion.

\section{Instanton generated interaction}\label{section:vertex}

Our calculation for TSSA is based on the presence of the intrinsic spin-flip during the quark-gluon interaction already on the quark level.
The generating functional for such non-perturbative interaction was obtained previously\cite{Kochelev:1996pv}. Later it was generalized in order to preserve the chiral invariance\cite{Diakonov:2002fq}. The generalized interaction Lagrangian has form
\begin{equation}\label{eq:Lagrangian}
    \mathcal{L}_I=-i g_s \frac{\mu_a}{4 m_q} \bar{\psi} \, t^a [\sigma_{\mu\nu} e^{i\gamma_5 \vec{\tau}\cdot \vec{\phi}/F_{\pi}}] \psi \, G^a_{\mu\nu},
\end{equation}
where $g_s$ is the strong coupling constant, $\mu_a$ is the anomalous quark chromomagnetic moment(AQCM), $m_q$ is the constituent quark mass, $t^a$ are $SU(3)$  color matrices,  $\sigma_{\mu \nu}=\frac{1}{2}[\gamma_\mu , \gamma_\nu]$.
$\vec{\tau}$ are Pauli matrices acting in the flavor space, 
$\vec{\phi}$ is the pion field, $F_\pi=93$ MeV is the pion decay constant. $G^a_{\mu\nu}$ is the gluon field strength.
This effective interaction is obtained by expanding t'~Hooft interaction in the power series in the gluon field strength, assuming a big spatial size of the gluon fluctuations.


Based on the Lagrangian (\ref{eq:Lagrangian}), the full interaction vertex is
\begin{equation}\label{eq:int_vertex}
U_\mu^a = i g_s t^a \left( \gamma_\mu - \sigma_{\mu \nu} q_\nu F(k_1,k_2,q) e^{i \gamma_5 \vec{\tau} \cdot \vec{\phi} / F_{\pi} } \right).
\end{equation}
The first term $\gamma_\mu$ corresponds to usual pQCD interaction. The second term is from effective low-energy action  Eq.(\ref{eq:Lagrangian}). 
$k_{1,2}$ are the momenta of incoming and outgoing quarks, $q=k_2-k_1$. 
The form factor $F$ is calculated in the instanton liquid model\cite{Schafer:1996wv,Diakonov:2002fq}:
\begin{equation}
\begin{split}
F(k_1,k_2,q)&=
\frac{\mu_a}{2m_q} \Phi_q 
\left(\frac{|k_1|\rho_c}{2} \right) \Phi_q 
\left(\frac{|k_2|\rho_c}{2} \right)
F_g(|q|\rho_c),
\\
\Phi_q(z)&=-z\frac{d}{dz}(I_0(z)K_0(z)-I_1(z)K_1(z)),
\\
F_g(z)&=\frac{4}{z^2}-2K_2(z),
\end{split}
\end{equation}
where are the $I_{\nu}(z)$ and $K_{\nu}(z)$ are the modified
Bessel functions. 
$\rho_c \approx 1.67$ GeV$^{-1}$ ($1/3$~fm) is the average instanton size.
In our calculations all quarks are on mass shell, therefore $\Phi_q=1$ and we will omit it further.

The AQCM $\mu_a$ is calculated in the framework of the instanton liquid model\cite{Kochelev:1996pv} is
\begin{equation}\label{eq:AQCM_value_inst_model}
\mu_a=-\frac{3\pi (m_q\rho_c)^2}{4\alpha_s(\rho_c)}.
\end{equation}
AQCM in pQCD appears at higher order $\alpha_s$ corrections. Therefore, it has a small value $\mu_{pQCD}=\alpha_s/2\pi \approx 10^{-2}$.
In contrast, the instanton generated AQCM is of the order of $1$.
Moreover, instanton liquid model gives the sign of AQCM and, in its turn, determines the sign of observed TSSA. 
Eq.~\ref{eq:AQCM_value_inst_model} is obtained in the massless chiral limit. One should not be confused that the $\mu_a$ increases with the quark mass. $m_q$ is the constituent mass and this equation could not be applied for heavy $c, b$ and $t$ quarks.


If we expand the exponent in Eq.(\ref{eq:int_vertex}) into series and cut it on
the second term, we get three types of vertices: traditional perturbative, chromomagnetic and the vertex with pion,
\begin{equation}\label{eq:int_vertex_expanded}
\begin{split}
U_\mu^a = i g_s t^a \Bigg( &\gamma_\mu - \sigma_{\mu \nu} q_\nu F(k_1,k_2,q) \\
& - i \frac{\vec{\tau} \vec{\phi}}{F_{\pi}} \gamma_5 \sigma_{\mu \nu} q_\nu F(k_1,k_2,q) \Bigg).
\end{split}
\end{equation}
We neglect the higher order terms. Their contribution to cross section is expected to be suppressed in the large $N_c$ limit by factor $1/N_c$ because $F_{\pi}\sim \sqrt{N_c}$\cite{Witten:1979kh}.
Moreover, due to increasing of the final particles number, it should be suppressed at large $x_F$ where TSSA is observed.

\section{Calculation of cross section}\label{section:Xsec}

\begin{figure}
\centering
  \includegraphics[width=0.9 \columnwidth]{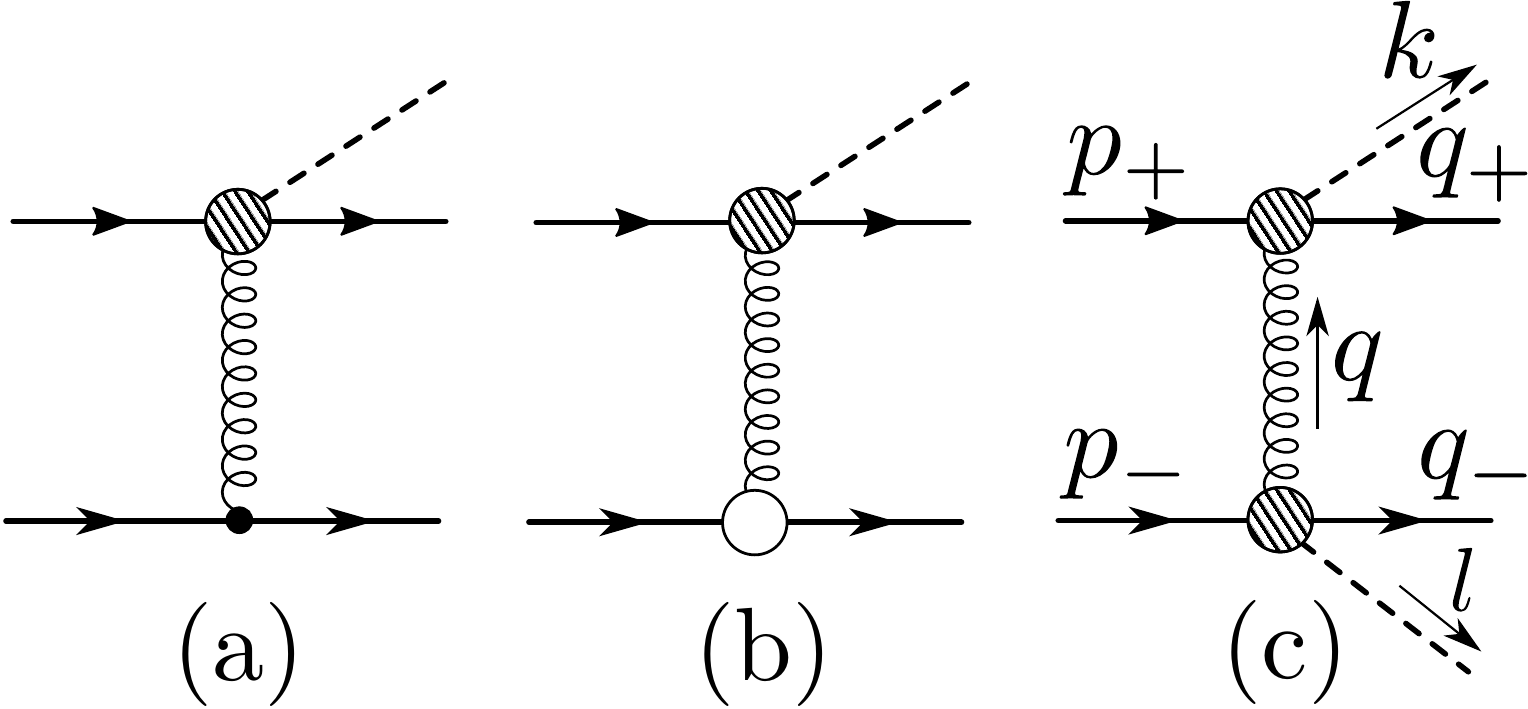}
\caption{Contributions to the pion production cross section and notation for momenta. The small dot denotes the perturbative vertex. The white blob is for the instanton induced interaction, it corresponds to the second term in Eq.(\ref{eq:int_vertex_expanded}). The shaded blob corresponds to the last term in Eq.(\ref{eq:int_vertex_expanded}) with the pion.}
\label{fig:Xsection_contributions}
\end{figure}

We are interested in process  $p^\uparrow p\to\pi X$.
Three parton subprocesses give contribution to the cross section. 
They are shown on Fig.\ref{fig:Xsection_contributions}. 
The diagram (a) was calculated before in \cite{Kochelev:2015pha} using an  assumption that the pion fragmentates in the same kinematic region as the quark $q_{+}$, i.e. the pion and quark flight approximately in the same direction.
In the present work we implement more rigorous calculation for the phase space and calculate additional contributions shown on panels (b) and (c) of Fig.\ref{fig:Xsection_contributions}.
The contribution (b) has the chromomagnetic vertex on the bottom quark line instead of the perturbative one. 
The diagram (c) $q+q\to 2\pi+2q$ is essentially different. 
In our model we have the pion directly in the interaction vertex and should consider the process where the pion is inside of an unobserved inclusive state $X$.
As the first step we study the partonic cross section and its features.
Then we calculate the hadron cross section as convolution of partonic one with parton densities.

\subsection{Parton cross section}

In massless limit the parton cross section is
\begin{equation}\label{eq:parton_Xsec}
  d\hat{\sigma}=\frac{|\mathcal{M}|^2}{2\hat{s}}d\hat{R}_i,
\end{equation}
where $d\hat{R}_i$ is the phase space for $i$ number of particles. 
In our case it can be three $d\hat{R}_3$ and four $d\hat{R}_4$. 
We use the hat symbol to emphasize that the phase space is expressed in terms of momenta and energies calculated in the \emph{parton} c.m. frame. 
This frame moves in respect to the hadron c.m. frame.
$\hat{s}$ is the total energy of colliding partons.


In calculation we use the following Sudakov decomposition for momentum vectors:
\begin{equation}\label{eq:momenta_decomposition}
\begin{split}
  k   &= x p_+ + \beta_k p_- + k_\perp,  \\
  q_+ &= \alpha_+ p_+ +\beta_+ p_- + q_{+ \perp},  \\
  q_- &= \alpha_- p_+ +\beta_- p_- + q_{- \perp}, \\
  q   &= \alpha p_+ +\beta p_- + q_{\perp}, \\
  l   &= \alpha_l p_+ + z p_- + l_\perp.
\end{split}
\end{equation}
$x$ and $z$ are parts of longitudinal momentum of the initial quarks carried by $k$ and $l$ pions correspondingly. 
$p_+$ and $p_-$ are light-cone vectors:
\begin{align}
    p_+&=(\sqrt{\hat{s}}/2,\sqrt{\hat{s}}/2,0_{\perp}),& \quad p_- &= (\sqrt{\hat{s}}/2,-\sqrt{\hat{s}}/2,0_{\perp}), \nonumber \\
    \hat{s} &= (p_+ + p_-)^2,& \quad p_+^2 &= p_-^2=0.
\end{align}

Using this momenta decomposition the phase space $d\hat{R}_3$ becomes
\begin{align}
  d\hat{R}_3 =
\frac{1}{4(2\pi)^5} \frac{dx \, d^2\!k_{\perp} \, d^2\!q_{\perp}}{x (1-x) \hat{s}}.
\end{align}
Integration over the transverse transferred momenta $q_{\perp}$ can be transformed to integration over the invariant mass $M_k^2=(k+q_{+})^2$: 
\begin{equation}\label{eq:dR3}
  d\hat{R}_3= \frac{1}{2^7 \pi^5} \frac{dx \, d^2\!k_{\perp}}{x^2 \hat{s}}\int_0^{E^2_{\text{sph}}} \!\!\!\!\!\!dM_k^2 \int_0^\pi \!\!\!\!\! d\tilde{\phi},
\end{equation}
where $E_{\rm sph}$ is the sphaleron energy, $\tilde{\phi}$ is the azimuthal angle of an auxiliary vector $\tilde{q}_\perp=x q_\perp - k_\perp$(see Appendix~\ref{appendix:PS} for details).

Sphaleron energy $E_{\rm sph} = \frac{3\pi}{2\rho_c}$\cite{Zahed:2002sy,Diakonov:2002fq} determines the height of potential barriers between different topological vacuums.
Instanton describes tunneling through that barrier, therefore the instanton induced vertex works only at energies less than the height of the barrier.

For diagram Fig.~\ref{fig:Xsection_contributions} (c) we need the 4-particle phase space. Using Sudakov decomposition \ref{eq:momenta_decomposition} it is
\begin{equation}
  d\hat{R}_4 =
  \frac{dx dz d^2 \! k_\perp \, d^2 \! l_\perp \, d^2\! q_\perp}{8 (2\pi)^8 \hat{s} x (1-x) z (1-z)}.
\end{equation}
Similar to $d\hat{R}_3$, we change integration over transverse momenta to integration over the invariant mass $M_{l}^2=(l+q_{-})^2$, $d^2l_\perp \to d M_{l}^2 d\phi$. Notice that here we replace  $d^2l_\perp$, not $d^2q_\perp$,
\begin{equation}
  d\hat{R}_4 = \frac{dx dz d^2 \! k_\perp \, d^2\! q_\perp \, dM^2_{l}}{2^{11} \pi^7 \hat{s} x (1-x)}.
\end{equation}

Next step is calculation transition amplitudes $\mathcal{M}_{(a,b,c)}$. A letter corresponds to a panel on Fig.~\ref{fig:Xsection_contributions}. The amplitude for first diagram Fig.\ref{fig:Xsection_contributions}(a) is
\begin{equation}
\begin{split}
   |\mathcal{M}_{(a)}|^2 =&  \sum_{f,s,c} g_s \frac{C(q^2)}{F_\pi} (\bar{u}_{q_+} 
   \sigma_{\mu \lambda} q_\lambda  \gamma_5 t^a 
   u_{p_+}) 
   \\
  & \times (\bar{u}_{q_-} i \gamma_\nu t^{a'} u_{p_-}) D_{\mu\nu}^{aa'}(q^2) \times \Big[ h.c. \Big],
\end{split}
\end{equation}
where $ \sum_{f,s,c}$ short-notes averaging over spin, color and flavor summation for corresponded pion($\pi^0$, $\pi^{\pm}$). $D_{\mu\nu}^{aa'}(q^2)=-i g_{\mu\nu}\delta^{aa'}/q^2$ 
is the gluon propagator. 
$C(q^2)$ can be thought as an effective coupling:
\begin{equation}
  C(q^2) = g_s \frac{\mu_a}{2 m_q} F_g(q^2)
  = - \frac{3 \pi^{3/2} \rho_c^2 m_q}{4 \sqrt{\alpha_s(\rho_c)}}     F_g(q^2).
\end{equation}
Note that $\alpha_s$ in the nonperturbative vertex is taken at the instanton size scale.
This is the reason why we keep one $g_s$ inside of $C(q^2)$ and another, from perturbative vertex, outside. They supposed to be taken at different scales. Further, we will omit writing $q$-dependency of $C$ for shortness.


We are interested in forward scattering.
At such kinematics, for simplicity of calculation, we use Gribov's decomposition for $g_{\mu\nu}$ in the gluon propagator.
\begin{equation}
g_{\mu\nu}=\frac{2 p_{+\mu} p_{-\nu}}{\hat{s}}+\frac{2 p_{+\nu} p_{-\mu}}{\hat{s}}+g^{\perp}_{\mu\nu} \approx \frac{2 p_{+\mu} p_{-\nu}}{\hat{s}}.
\end{equation}
Such decomposition allows us to isolate the leading contributions to an  amplitude in the power of $\hat{s}$ and factorize fermion traces.
Using it we get for the amplitude (see Appendix \ref{appendix:amplitudes})
\begin{align}
|\mathcal{M}_{(a)}|^2&=
   \sum_{f} \frac{8}{9} g_s^2 \frac{C^2}{F_{\pi}^2} \frac{\hat{s}^2 (1-x)}{q_{\perp}^2}.
\end{align}
We keep the sum over flavor to indicate that expressions for $\pi^\pm$, $\pi^0$ are different.


In the case of the diagram Fig.\ref{fig:Xsection_contributions}(b), the difference is only in the trace over the bottom line,
\begin{align}\label{eq:M1_amplitude}
  |\mathcal{M}_{(b)}|^2 &= \sum_{f,s,c} \frac{C^2}{F_\pi} (\bar{u}_{q_+} 
  \sigma_{\mu \lambda} q_\lambda  \gamma_5 t^a 
  u_{p_+}) \nonumber \\
  & \times (\bar{u}_{q_-} i \sigma_{\nu \rho} q_\rho t^{a'} u_{p_-}) 
  D_{\mu\nu}^{aa'}(q) \times \Big[ h.c. \Big]
  \\
 & = \sum_{f} \frac{8}{9} \frac{C^4}{F_{\pi}^2} \hat{s}^2 (1-x). \nonumber
\end{align}
Notice that now the amplitude is proportional to $C^2$, not $g_s C$. 


The amplitude for the two pion contribution $q + q \to 2\pi + 2q$ Fig.~\ref{fig:Xsection_contributions}(c) is very similar to the case with one pion vertex. 
Now, the trace over the bottom fermion line is similar to the upper one.
\begin{equation}
\begin{split}
  |\mathcal{M}_{(c)}|^2 &=
  \sum_{f,s,c}  -\frac{C^2}{F_{\pi}^2} 
  (\bar{u}_{q_+} \sigma_{\mu \lambda} q_\lambda  \gamma_5 t^a u_{p_+}) \\
  &\times (\bar{u}_{q_-} \sigma_{\nu\rho} q_{\rho} \gamma_5 t^{a'} u_{p_-}) 
  D_{\mu\nu}^{aa'} \times \Big[ h.c. \Big] \\
  &=\sum_{f} \frac{8 C^4}{9 F_{\pi}^4} \hat{s}^2 (1-x)(1-z).
\end{split}
\end{equation}
Final formulas for contributions to the parton cross section shown on Fig.\ref{fig:Xsection_contributions} are 
\begin{align}\label{eq:partoc_Xsec_abc}
  d\hat{\sigma}_{(a)} &= \sum_{f}\int\limits_{0}\limits^{E^2_{\rm sph}} \!\!\! dM_k^2 \!\! \int\limits_{0}\limits^{\pi} \!\!\! d\tilde{\phi} \, \frac{g_s^2 C^2}{9(2\pi)^5 F_{\pi}^2} 
  \frac{1-x}{q_\perp^2 x^2}dx d^2k_\perp,
  \\
  d\hat{\sigma}_{(b)} &=\sum_{f} \int\limits_{0}\limits^{E^2_{\rm sph}} \!\!\! dM_k^2 \!\! \int\limits_{0}\limits^{\pi} \!\!\! d\tilde{\phi} \, \frac{ C^4}{9(2\pi)^5F_{\pi}^2} 
  \frac{1-x}{x^2}dx d^2k_\perp,
  \\
  d\hat{\sigma}_{(c)} &= \sum_{f} \!\!
  \int\limits_{0}\limits^{E^2_{\rm sph}} \!\!\! dM_k^2 \!\! \int\limits_{0}\limits^{\pi} \!\!\! d\tilde{\phi} \, 
  \frac{C^4 E^2_{\text{sph}}}{9 \, 2^{10} \pi^7 F_{\pi}^4} 
  \frac{(1-x)}{x^2} \, dx \, d^2k_\perp.
\end{align}
The detailed derivation of this equations is given in Appendix \ref{appendix:amplitudes}.

\subsection{\texorpdfstring{$pp\to\pi X$}{pp to pi X} cross-section}
The next step is to calculate observables on the hadron level. 
Differential hadron cross-section is a convolution of parton distribution functions(PDF) and the parton cross section
%
%
%
\begin{equation}
E_k \frac{d \sigma}{d^3k} = \sum_f \int^{x_a^{\max}}_{x_a^{\min}} \!\!\!\!\!\! dx_a \int^{x_b^{\max}}_{x_b^{\min}} \!\!\!\!\!\! dx_b \, f(x_a) f(x_b) \frac{2 E_k}{\sqrt{s}x_a} \frac{d \hat\sigma}{dx d^2 k_{\perp}}.
\end{equation}
The flavor sum $\sum_f$ indicates the proper summation for a corresponding pion.
The explicit formula for the $\pi^0$ production cross section is
\begin{equation}\label{eq:hadron_xsec_final}
\begin{split}
E_{k} \frac{d \sigma}{d^3k} =  &3\iint \! dx_a  dx_b \Big( f_u(x_a)+f_d(x_a) \Big)  \\
\times & \Big(f_u(x_b)+f_d(x_b)\Big) \frac{\sqrt{x_T^2 + x_F^2}}{x_a} \frac{d \hat\sigma_{(c)}}{dx d^2 k_{\perp}} \\
& + \iint \! dx_a  dx_b \Big( f_u(x_a)+f_d(x_a) \Big) \\
& \times \Big(f_u(x_b)+f_d(x_b)\Big) \frac{\sqrt{x_T^2 + x_F^2}}{x_a} \frac{(d \hat\sigma_{(a)}+d \hat\sigma_{(b)})}{dx d^2 k_{\perp}}.
\end{split}
\end{equation}
Factor $3$ in the first line is the result of summation over unobserved  pions($\pi^\pm , \pi^0$) in inclusive state $X$, produced from the bottom vertex Fig.\ref{fig:Xsection_contributions}(c).
In the case $\pi^+$ production the cross section is
\begin{equation}
\begin{split}
E_{k} \frac{d \sigma}{d^3k} = & 6 \iint \! dx_a  dx_b f_u(x_a) \Big(f_u(x_b)+f_d(x_b)\Big) \\
&\times \frac{\sqrt{x_T^2 + x_F^2}}{x_a} \frac{d \hat\sigma_{(c)}}{dx d^2 k_{\perp}} \\
&+ 2 \iint \! dx_a  dx_b f_u(x_a) \Big(f_u(x_b)+f_d(x_b)\Big)  \\
&\times \frac{\sqrt{x_T^2 + x_F^2}}{x_a} \frac{(d \hat\sigma_{(a)}+d \hat\sigma_{(b)})}{dx d^2 k_{\perp}}.
\end{split}
\end{equation}
$\pi^-$ cross section is given by replacing $f_u(x_a) \to f_d(x_a)$.

In order to determine integration limits for $x_{a,b}$, notice that one could reduce  $2\to 3$ and $2\to 4$ parton subprocesses to the $2 \to 2$ case if combines all particles except the detected pion into an effective particle with the mass square $X^2$. 
We could not neglect this invariant mass since it is of order of $s$.
From
\begin{equation}
\hat{s}+\hat{t}+\hat{u}=X^2,
\end{equation}
one could relate $x_a$ and $x_b$.
Using $X^2 \geq 0$, maximum and minimum values for $x_a$ and $x_b$ are(see Appendix~\ref{appendix:int_limits}): 
\begin{align}\label{eq:xa_xb_limits}
&x_a^{\min} = \frac{4 x_F^2}{4x_F - x_T^2};  &x_a^{\max}=1;\\
&x_b^{\min} = \frac{k^2_{\perp}/x}{x_a(1-x)s};  &x_b^{\max}=1,
\end{align}
where $x=x_F/x_a$. 

\section{Single-spin asymmetry}\label{section:ssa}
Consider scattering of the proton with transverse polarization vector $\vec{a}$ and momentum $p_+$ and other unpolarized proton with momentum $p_-$. In semi-inclusive process the pion with momentum $k$ is produced.

\begin{figure}
	\includegraphics[width=0.7 \columnwidth]{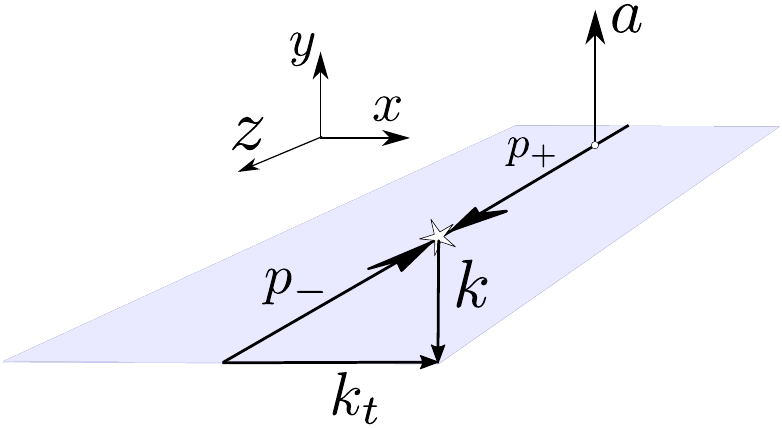}
	\caption{Kinematics of TSSA. The polarized proton with the momentum $p_{+}$ moves in $+z$. The polarization vector $a$ is in $+y$ or $-y$ direction. The pion momentum lies in $zx$ plane.}
	\label{fig:kinematics_ssa}
\end{figure}

For the TSSA calculation it is crucial to define a coordinate system, because the sign of TSSA depends on it. 
We choose the standard right-hand coordinate system. 
The initial polarized proton moves in $+z$ direction and its polarization vector is along $y$ axis, Fig.\ref{fig:kinematics_ssa}.
Positive TSSA means that more pions produced in $+x$ half-space when the proton has spin in $+y$ direction.


Transverse single spin asymmetry(or analyzing power) is defined as
\begin{equation}\label{eq:ssa_definition}
A_N=\frac{d\sigma_{\uparrow} - d\sigma_{\downarrow}}
{d\sigma_{\uparrow} + d\sigma_{\downarrow}} = \frac{d\Delta\sigma}{ 2 d\sigma}.
\end{equation}
Arrows $\uparrow$ and $\downarrow$ denote the spin polarization vector of the proton in $+y$ and $-y$ direction correspondingly.
We consider only tree-level diagrams for the unpolarized cross-section in the denominator of Eq.(\ref{eq:ssa_definition}). 
As it will be shown later, it is enough to reproduce cross section data. 
Moreover, we expect that higher orders are suppressed by instanton density and $\alpha_s$.

Polarized parton cross section is related with hadron cross section as a convolution with polarized PDFs.
\begin{equation}
d\sigma_{\uparrow} = \sum_{f} \iint dx_a dx_b \, f_{a^\uparrow /A^\uparrow}(x_a) f(x_b) d\hat{\sigma}_{\uparrow} 
\end{equation}
\begin{equation}
  \Delta\sigma = \Delta_T f_{a/A} \otimes f_b \otimes \Delta \hat{\sigma}.
\end{equation}
$\Delta_T f_{a/A}$ is the transversity distribution -- the difference between the probabilities to find parton $a$ polarized parallel and anti-parallel to the polarization of hadron $A$. 

Transverse polarization state can be represented as superposition of helicity states: 
\begin{equation}\label{eq:spin_decomposition}
|\!\!\uparrow\downarrow \rangle = \frac{1}{\sqrt{2}} (| + \rangle \pm i | - \rangle).
\end{equation}
Using this we can rewrite the difference of amplitudes with opposite transverse polarizations as a product of helicity amplitudes:
\begin{equation}\label{eq:helicity_TSSA}
|\mathcal{M}_{\uparrow}|^2-|\mathcal{M}_{\downarrow}|^2 
= 2 \Im (\mathcal{M}_{+} \mathcal{M}^*_{-}). 
\end{equation}
$\pm$ mean helicity of initial parton in the polarized proton. We sum over polarization of other particles.
\begin{figure}
\begin{center}
\includegraphics[width=\columnwidth]{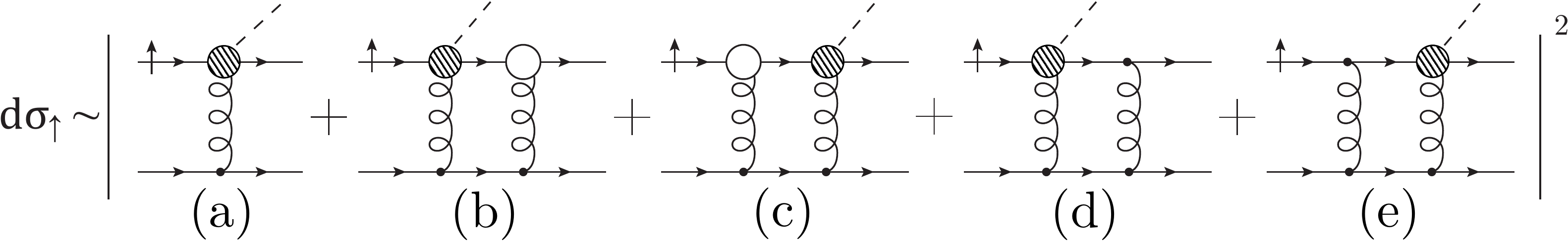}
\end{center}
\caption{The set of considered diagrams that give contributions to the total scattering amplitude. Notation for interaction vertices is similar to Fig.~\ref{fig:Xsection_contributions}.}
\label{fig:all_diagrams_in_TSSA}
\end{figure}
$\mathcal{M}_{\pm}$ has five parts shown on  Fig.\ref{fig:all_diagrams_in_TSSA}. Until now both $\mathcal{M}_+$ and $\mathcal{M}_-$ contain spin-flip and non-flip amplitudes. But only the interference between spin-flip (a,d,e) and non-flip(b,c) diagrams survives in TSSA.
In this light, one could think about $\mathcal{M}_+ \mathcal{M}^*_-$ as a product of spin-flip and non-flip amplitudes. 
Leading contribution into $\Delta\sigma$ comes from interference between (a) and (b+c) diagrams. 
We expect that the interference between (b+c) and (d+e) diagrams is suppressed due to additional $\alpha_s$. Moreover, because they have the same structure, phase shift between them is small. 

Upper line should have an odd number of chromomagnetic vertices and the bottom line -- an even number or all perturbative. 
Firstly we look the case with all perturbative vertices on the bottom line.
\begin{figure}
\begin{center}
    \includegraphics[width=\columnwidth]{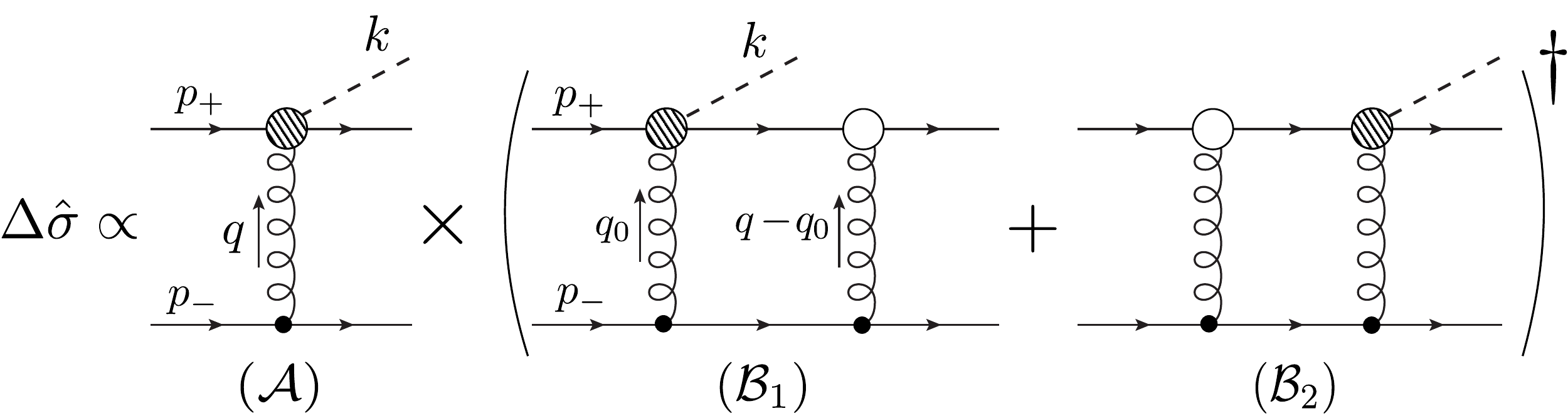}
    \caption{The leading contribution to TSSA. The left loop diagram we denote as $\mathcal{B}_1$, the right one as $\mathcal{B}_2$. The tree-level diagram is $\mathcal{A}$.}
    \label{fig:delta_sigma_leading_interf}
\end{center}
\end{figure}

We use the momentum notation that is shown on Fig.\ref{fig:delta_sigma_leading_interf}.
Sudakov's decomposition for momentum vectors is as before, Eq.(\ref{eq:momenta_decomposition}). 
A new vector $q_0$ is decomposed as:
\begin{align}
    q_0&=\alpha_0 p_+ + \beta_0 p_- + q_{0\perp}.
\end{align}
$\Delta \hat{\sigma}$ is proportional to the interference of spin flip and non flip diagrams Eq.(\ref{eq:helicity_TSSA})
\begin{equation}
    \Delta\hat{\sigma} = \frac{1}{2\hat{s}} 2 \cdot 2 \overline{\sum_{s,c}} \Im[\mathcal{A_+} (\mathcal{B}_{1-}+\mathcal{B}_{2-})^*]d\hat{R}_3.
\end{equation}
Factor $1/2\hat{s}$ is the flux of initial particles. In the numerator first factor 2 appears because 
$
\mathcal{A}_{+} (\mathcal{B}_{1-} + \mathcal{B}_{2-})^*=\mathcal{A}^*_{-} (\mathcal{B}_{1+}+\mathcal{B}_{2+})
$
\cite{Dixon:2013uaa} and the second from Eq.(\ref{eq:helicity_TSSA}). $\overline{\sum}_{s,c}$ symbolically denotes averaging over spin and color states. Three-particles phase space $d\hat{R}_3$ was calculated before. 

Using Gribov decomposition for $g_{\mu\nu}$ we factorize diagrams to upper and lower parts.
The interference between first and second diagram on the Fig.\ref{fig:delta_sigma_leading_interf} is
\begin{equation}
\begin{split}
\mathcal{A}_{+} \mathcal{B}_{1-}^* =& \frac{1}{2\cdot9} 
\Big(
\frac{2}{\hat{s}}
\Big)^3 
g_s^3 \frac{C^3}{F_{\pi}^2} \int \! \! \frac{d^4q_0}{(2\pi)^4} 
\\
& \times \frac {\mathrm{Tr}[t^a t^b t^c] \mathrm{Tr}[t^{a'} t^{b'} t^{c'}] 
	\delta_{a a'}\delta_{b b'}\delta_{c c'} (U_1 D)}
{q^2 q_0^2 (q-q_0)^2 (p_+ + q_0 - k)^2 (p_{-} - q_0)^2}.
\end{split}
\end{equation}
$U_1$ and $D$ are products of gamma matrices corresponded upper and bottom fermion lines respectively. 
Factors 2 and 9 in denominator are from averaging over spins of the unpolarized quark and over color states. The color trace is
\begin{equation}
\mathrm{Tr}[t^a t^b t^c] \mathrm{Tr}[t^{a'} t^{b'} t^{c'}] \delta_{a a'}\delta_{b b'}\delta_{c c'}=-2/3.
\end{equation}
We calculate the imaginary part by putting fermions in the loop on mass shell. After collecting all $i$ and signs in vertices
\begin{align}
\Im (\mathcal{A B}_1^*) =& -\frac{2/3}{2\cdot 9}\Big(\frac{2}{\hat{s}}\Big)^3 g_s^3 \frac{C^3}{F_{\pi}^2} \int \! \! \frac{d^2 q_{0\perp} d\alpha_0 d\beta_0}{(2\pi)^4} \frac{\hat{s}(-2\pi i)^2}{2\cdot 2i}  \nonumber
\\
& \times \frac {\delta((p_- -q_0)^2) \delta((p_+ +q_0-k)^2) U_1 \, D}{q^2 q_0^2 (q_0-q)^2} \nonumber 
\\
=& -\frac{g_s^3}{54 \hat{s}^4 \pi^2}\frac{C^3}{F_{\pi}^2} \int \! \! \frac{d^2 q_{0\perp}}{(1-x)}\frac {U_1 \, D}{q_{\perp}^2 q_{0\perp}^2 (q_{0\perp}-q_{\perp})^2},
\label{eq:ImAB1_initial}
\end{align}
where $d^4q_0 = \frac{\hat{s}}{2}d\alpha_0 d\beta_0 d^2q_{0\perp}$ was used. Notice that the loop integral in $\mathcal{AB}_1^*$ is restricted by the sphaleron energy, similar to the phase space integral, $(p_+ + q_0)^2<E_{\textrm{sph}}^2$.

For the upper fermion line we have
\begin{align}
U_1 =&  \bar{u}_{p_+}(-) 
\cancel{q}_{0\perp} \cancel{p}_- \gamma_5
(\cancel{p}_+ + \cancel{q}_0 - \cancel{k})
(\cancel{q}_{\perp} - \cancel{q}_{0\perp}) \cancel{p}_- \nonumber
\\
&
\times (\cancel{p}_+ + \cancel{q} - \cancel{k})
\gamma_5  \cancel{p}_- \cancel{q}_{\perp}
u_{p_+}(+)  \nonumber
\\
=& -2 (1-x)^2 \hat{s}^3 (q_{0\perp}^2 q_x - q_{\perp}^2 q_{0 x}),
\end{align}
where subscript $x$ denotes the component of a vector along $x$-axis. $u(\pm)$ is the spinor for a quark in the corresponded helicity state. 
For the bottom quark line with all perturbative vertices the trace is
\begin{align}
D = \text{Tr}[(\cancel{p}_- -\cancel{q})\cancel{p}_+ \cancel{p}_- \cancel{p}_+ (\cancel{p}_--\cancel{q}_0) \cancel{p}_+]= 2 \hat{s}^3.
\end{align}

The second contribution shown on Fig.\ref{fig:delta_sigma_leading_interf} is given by the formula:
\begin{align}\label{eq:ImAB2}
\Im \mathcal{A B}_2^* =
 -\frac{g_s^3}{54 \hat{s}^4 \pi^2}\frac{C^3}{F_\pi^2}
\int \! \! d^2 q_{0\perp}\frac {U_2 \, D}{q^2_\perp q_{0\perp}^2 (q_{0\perp}-q_\perp)^2},
\end{align}
\begin{align}
U_2 =&  \bar{u}_{p_+}(-) 
\cancel{q}_{0\perp} \cancel{p}_- 
(\cancel{p}_+ + \cancel{q}_{0\perp})
(\cancel{q}_{\perp} - \cancel{q}_{0\perp}) \cancel{p}_- \gamma_5 \nonumber
\\  
&\times (\cancel{p}_+ + \cancel{q} - \cancel{k}) \gamma_5 \cancel{p}_- \cancel{q}_{\perp} 
u_{p_+}(+)  \nonumber
\\ 
=& 
2 (1-x) \hat{s}^3 (q_{0\perp}^2 q_x - q_{\perp}^2 q_{0 x}).
\end{align}

The absence of additional $(1-x)$ in the trace $U_2$ in comparison with $U_1$ is compensated by lack of $(1-x)$ in denominator of Eq.(\ref{eq:ImAB2}). 
The  trace $D$ is the same. Therefore, $\Im(\mathcal{AB}_1^*)$ and $\Im (\mathcal{AB}_2^*)$ differ by the sign and integration limits over $d^2q_{0\perp}$. 
Loop integral in $\mathcal{AB}_1$ is limited by the sphaleron energy. 
In contrast, the loop integral in $\mathcal{AB}_2$ does not have such limit.  Because the integrands are the same in an absolute value and with opposite sign, we can exclude part of the integration region where they are canceled out.
Nonzero contribution comes from region where $(p_+ + q_0)^2>E^2_{\textrm{sph}}$.

Combining this observations, the final result is
\begin{equation}\label{eq:TSSA_parton}
\begin{split}
  \frac{d\Delta \hat{\sigma}}{dx d^2k_{\perp}} 
  =& \frac{g_s^3}{27 \cdot 2^6 \pi^7}
  \frac{C^3}{F_\pi^2}
  \int_0^{E^2_{\textrm{sph}}}\!\!\!\!\!\!\!\!\!\!dM_{k}^2
  \!\!\int_{0}^{\pi} \!\!\!d\tilde{\phi}
  \int\!\!d^2q_{0\perp} 
  \\
  &\times \theta(M_0^2-E^2_{\textrm{sph}})
  \frac{(1-x)}{x^2} \frac{(q_{x}q_{0\perp}^2 -  q_{0x}q_{\perp}^2)} {q_{\perp}^2 q_{0\perp}^2 (q_{0\perp}-q_{\perp})^2},
\end{split}
\end{equation}
where $M_0^2=(p_+ + q_0)^2$.
\begin{figure}
	\includegraphics[width=.9\linewidth]{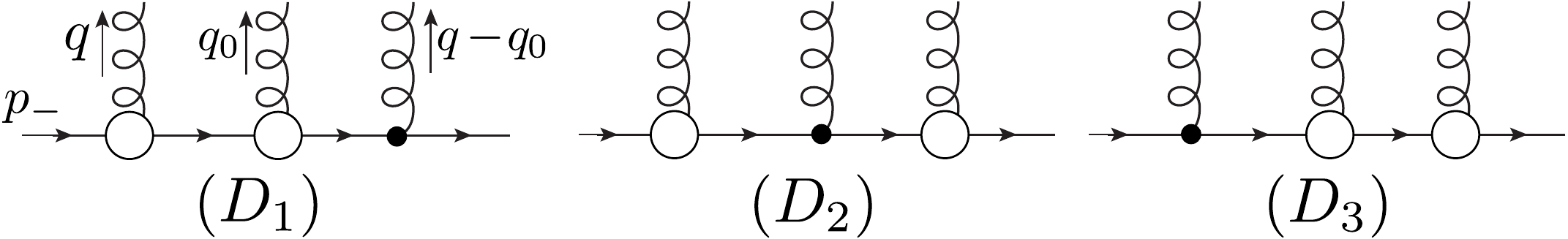}
	\caption{Additional contributions to TSSA from diagrams with chromomagnetic vertices on the bottom line.}
	\label{fig:CM_vertex_down_line}
\end{figure}

We also calculated contributions to TSSA from diagrams with chromomagnetic vertices on the bottom line (Fig.~\ref{fig:CM_vertex_down_line}).
There are three possible combinations which give for the trace:
\begin{align}
D_{1} & = 2 s^3 (q_{\perp} \cdot q_{0 \perp}), \\
D_{2} & = 2 s^3 (q^2 - (q_{\perp} \cdot q_{0 \perp})), 
\\
D_{3} & = 2 s^3 (q_0^2 - (q_{\perp} \cdot q_{0 \perp})). 
\end{align}
One should substitute this expressions instead of $D$ in Eq.\ref{eq:ImAB1_initial} and Eq.\ref{eq:ImAB2}, replacing accordingly couplings $g_s$ and $C$.


\section{Numerical results and discussion}\label{section:discussion}


For numerical estimations we use parameters provided by the instanton liquid model for QCD vacuum\cite{Schafer:1996wv,Diakonov:2002fq}. We choose  $F_{\pi}=93$~MeV, $m_q=90$~MeV, $\rho_c=1.6$~GeV$^{-1}(0.32~\textrm{fm})$.
It corresponds to AQCM with the value $\mu_a=-0.45$ and $\alpha(\rho_c)\approx 0.6$.
For perturbative coupling we use
\begin{equation}
  \alpha_s(q^2)=\frac{4\pi}{9 \ln (q^2/\Lambda_{\mathrm{QCD}}^2)} \theta(q^2-1/\rho_c^2),
\end{equation}
where $\Lambda_{\mathrm{QCD}}=200$~MeV. Choice of  $\Lambda_{\mathrm{QCD}}$ does not affect significantly numerical results. 
The step-function $\theta$ ``switches off" perturbative interaction at momenta lower than the instanton scale. 
It regularizes cross section, removing Landau pole and effectively works as a phenomenological gluon mass. 
Such procedure can be justified in terms of the potential between quarks.
In Cornell potential the linear term starts to dominate the Coulomb-like term from one gluon exchange at distances more than $0.3$~fm. 

\begin{figure}[tb]
\begin{minipage}{.49\columnwidth}
  \centering
  \includegraphics[width=\columnwidth]{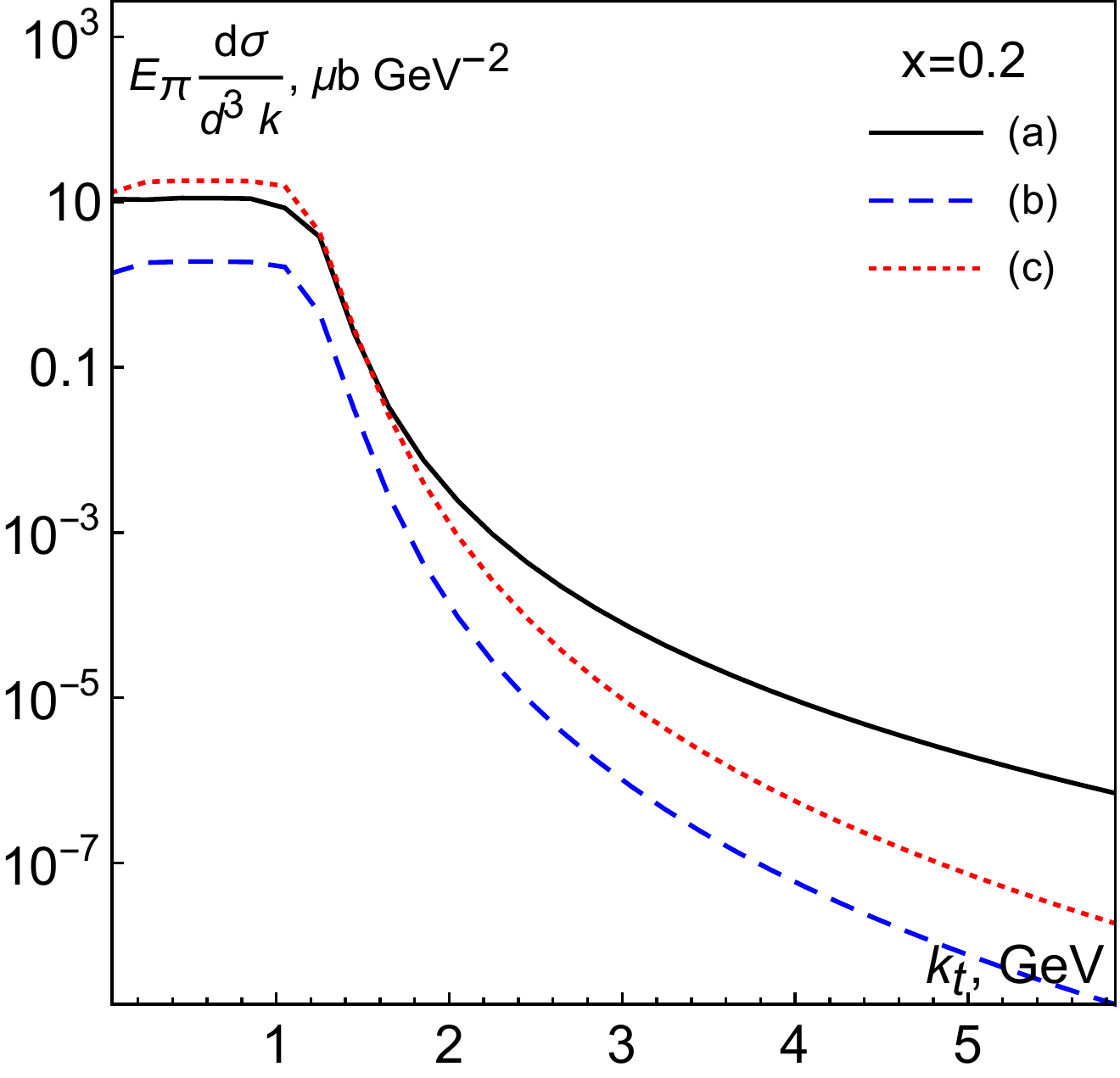}
  (a)
\end{minipage}%
\hspace{0.0em}
\begin{minipage}{.48\columnwidth}
  \centering
  \includegraphics[width=\columnwidth]{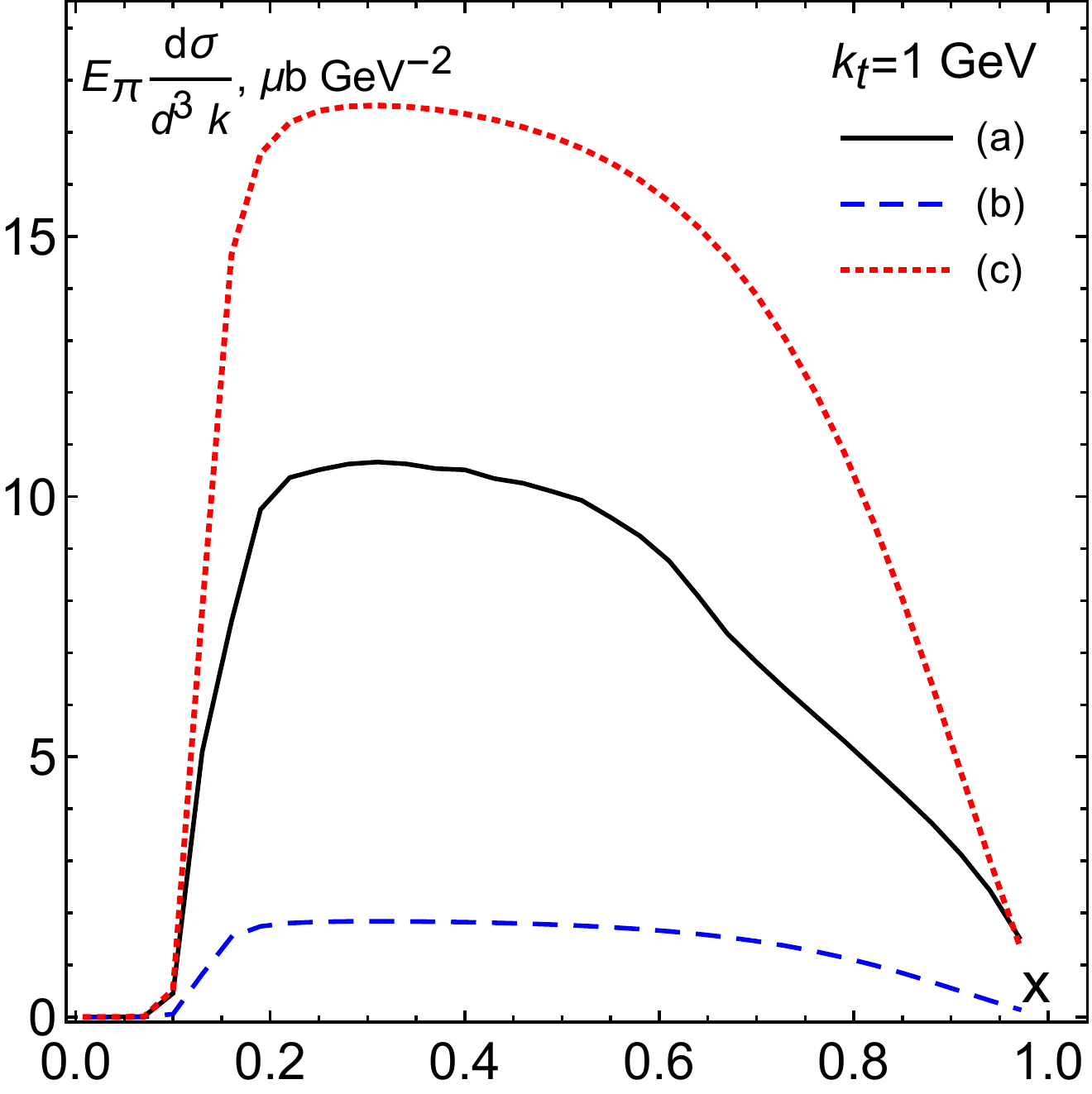}
  (d)
\end{minipage}\\
\begin{minipage}{.49\columnwidth}
  \centering
  \includegraphics[width=\columnwidth]{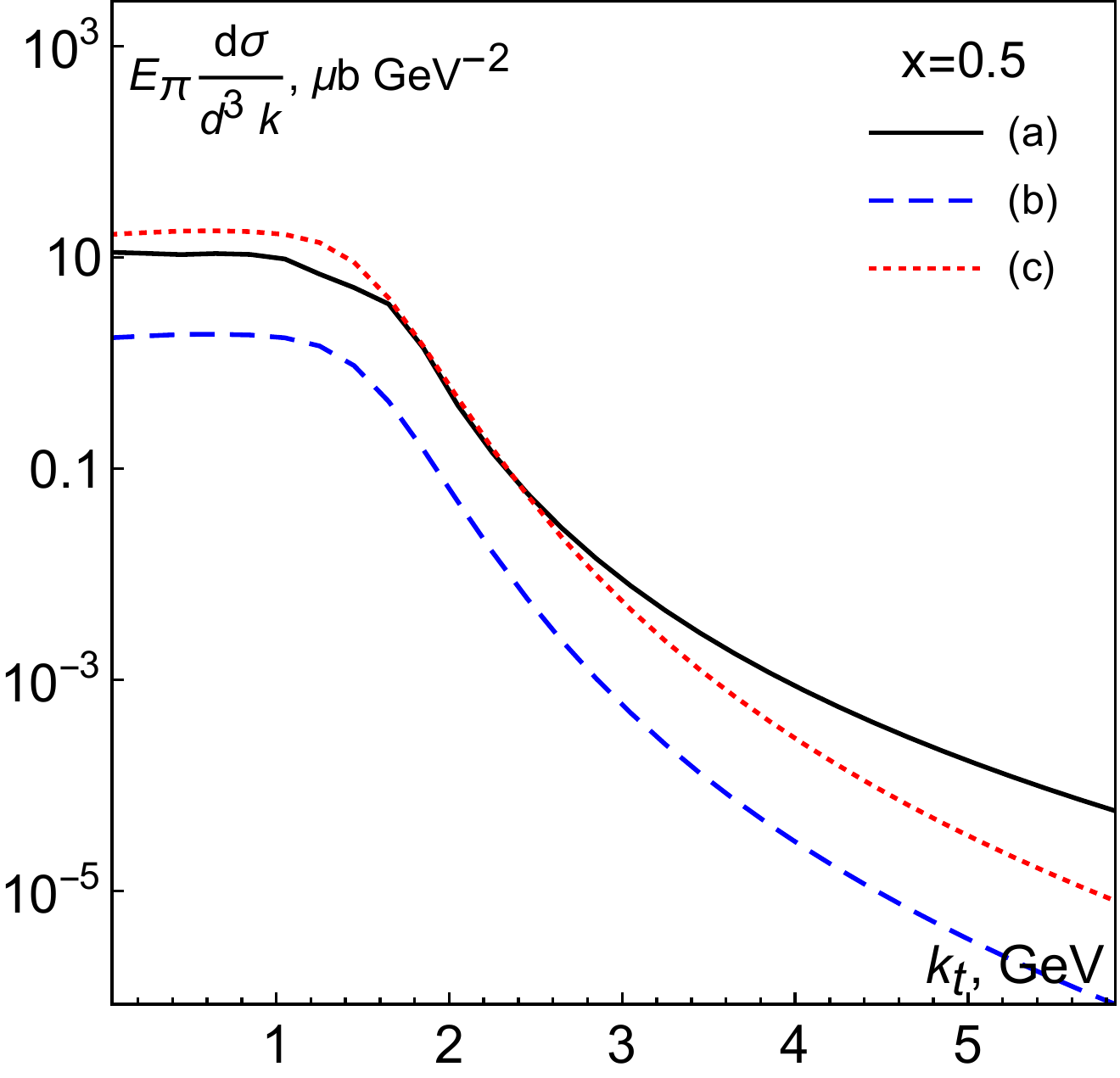}
  (b)
\end{minipage}
\hspace{0em}
\begin{minipage}{.48\columnwidth}
	\centering
	\includegraphics[width=\columnwidth]{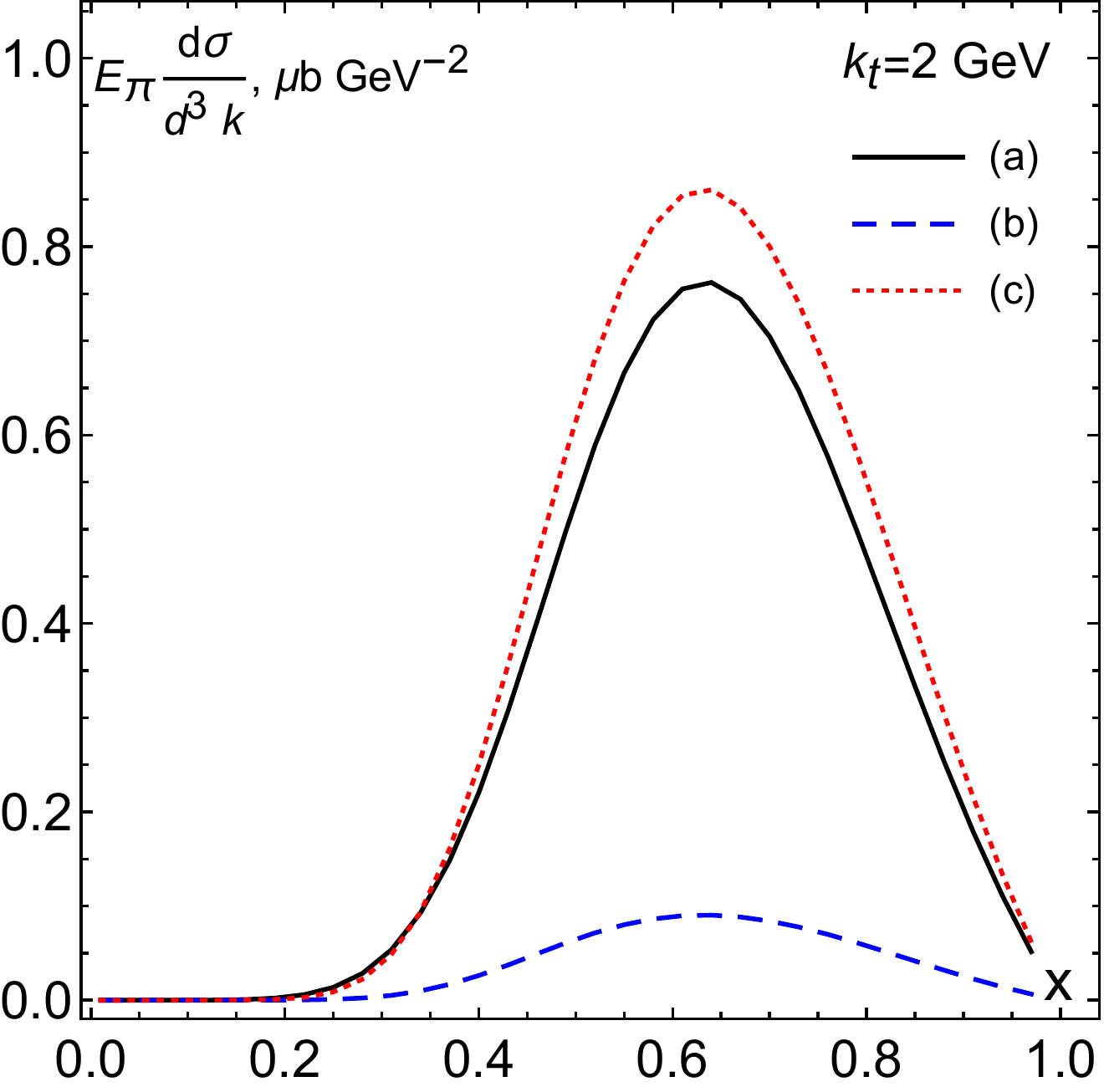}
	(e)
\end{minipage} \\
\begin{minipage}{.48\columnwidth}
	\centering
	\includegraphics[width=\columnwidth]{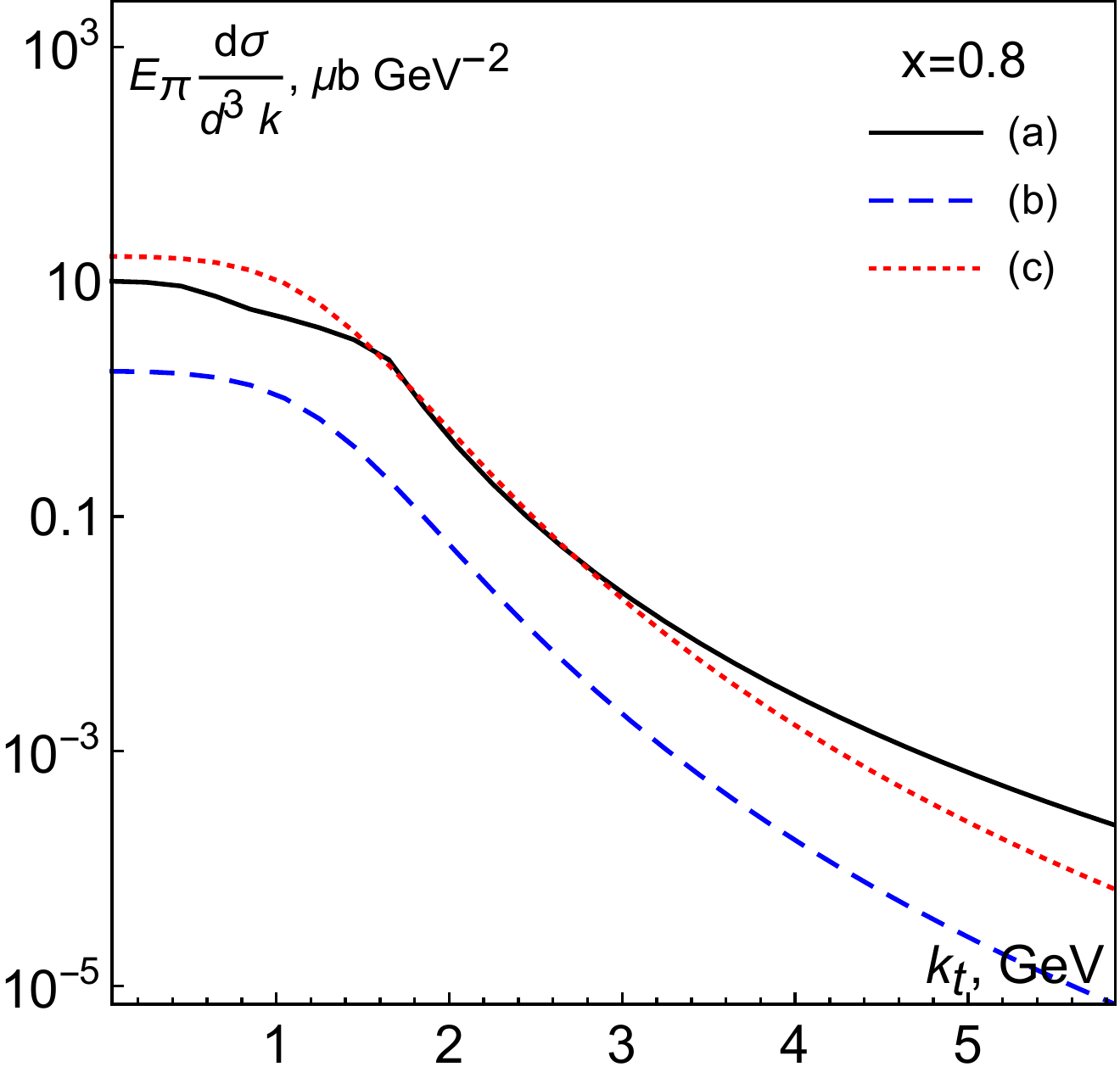}
	(c)
\end{minipage}
\hspace{0em}
\begin{minipage}{.49\columnwidth}
	\centering
	\includegraphics[width=\columnwidth]{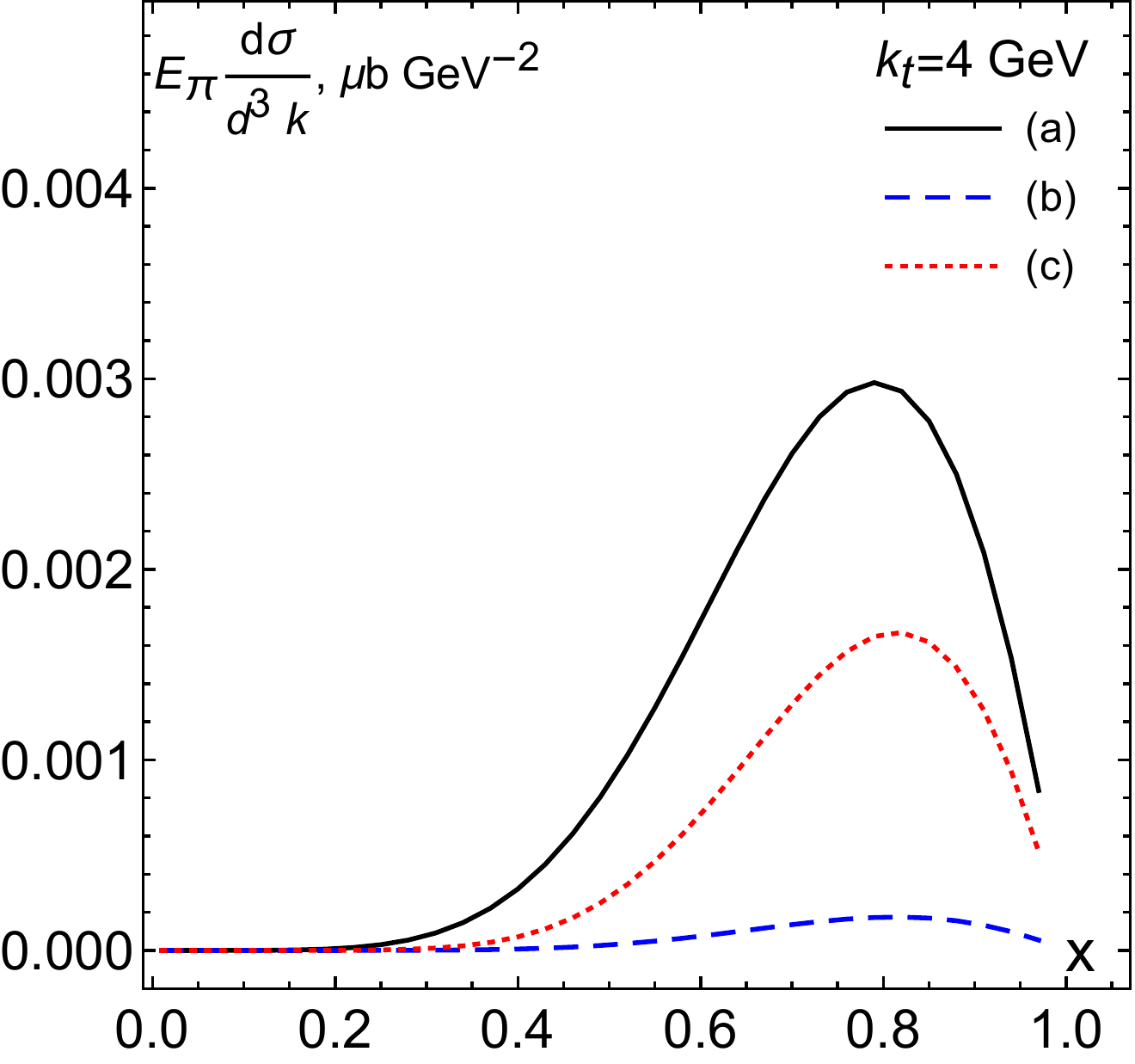}
	(f)
\end{minipage}
\caption{Differential cross section $qq \to \pi^0 X$  as function of $k_t$(left column) and $x$(right column). 
Solid line is for the contribution of Fig.\ref{fig:Xsection_contributions}(a). Dashed line is Fig.\ref{fig:Xsection_contributions}(b). The dotted line is for the two pion process Fig.\ref{fig:Xsection_contributions}(c). Parameters are as described in the text.}
\label{fig:parton_cs_vs_x_k}
\end{figure}

First, we will discuss results for parton cross section and TSSA in $qq\to \pi^0 X$ to demonstrate dynamics not affected by PDFs.
Further we use $k_t=k_{\perp}$.
The Fig.\ref{fig:parton_cs_vs_x_k} shows  contributions of different diagrams from Fig.\ref{fig:Xsection_contributions} to $\pi^0$ production cross section.
One could see that at chosen parameters contributions of diagrams (a) and (c) are of the same order while the contribution from (b) is smaller. The slope of cross-section with $k_t$ is determined by the shape of the form factor $F_g$.
All three contributions have similar dependency on $x$.
As expected, at high $k_t$ the diagram (a) with the perturbative vertex dominates.
  
$E_{\textrm{sph}}$ in Eq.~\ref{eq:partoc_Xsec_abc} determines minimal $k_t$ at which the whole quark-pion system has nonzero transverse momentum.
When $|k_t| > E_{\textrm{sph}}/2$, the exchanged gluon has to have  nonzero transverse momentum $q_\perp$ at any $x$. 
At $|k_t| \leq E_{\textrm{sph}}/2$, momenta $q_\perp$ can be zero and we get divergence. 
We avoid this by the cut of the perturbative coupling $\alpha_s$ described above.
This determines transition from ``flat'' behavior of the cross section at small $k_t<1.5$~GeV to falling.
%
\begin{figure}[tbh]
	\includegraphics[width=.48\columnwidth]{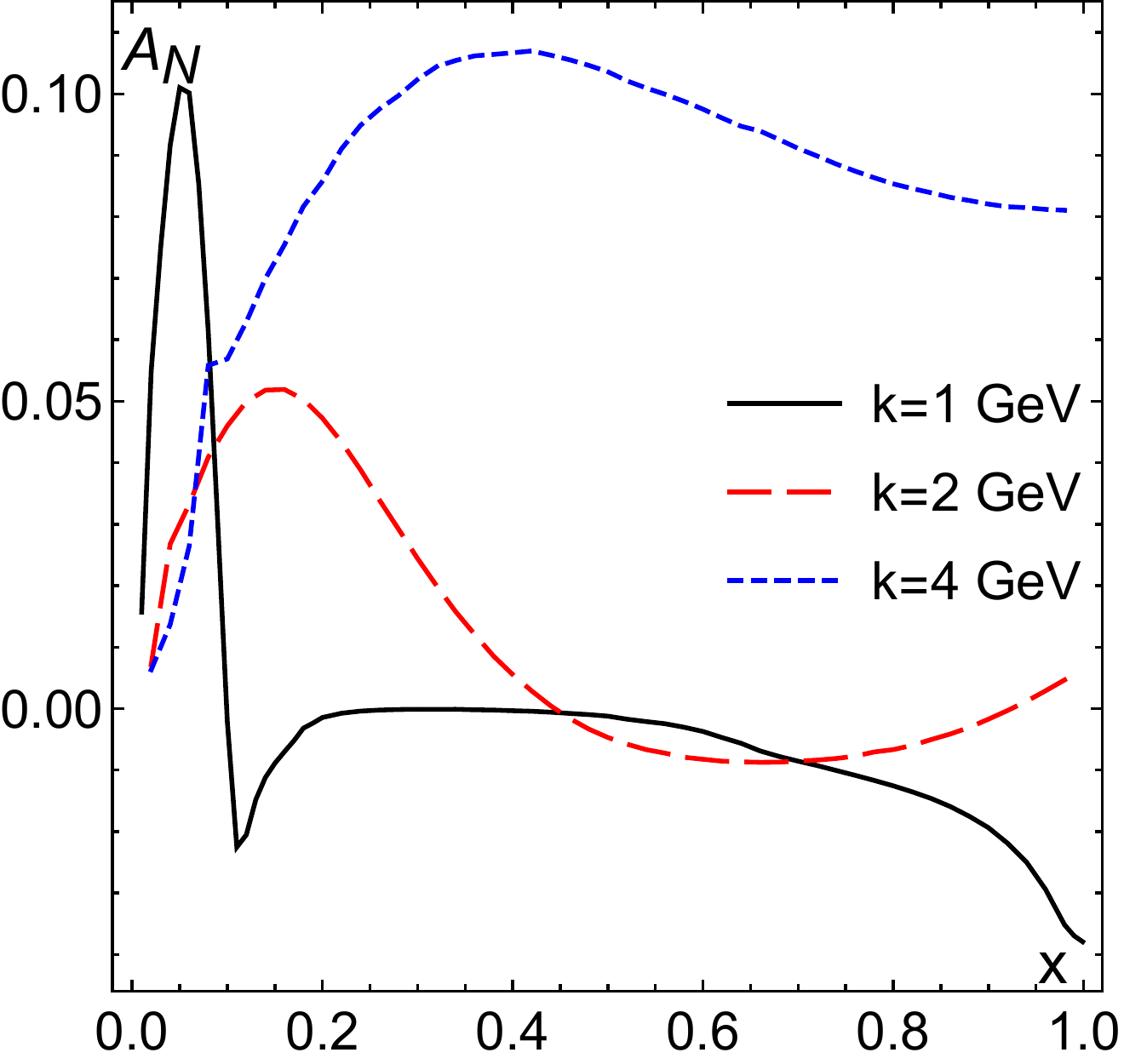}
	\includegraphics[width=.49\columnwidth]{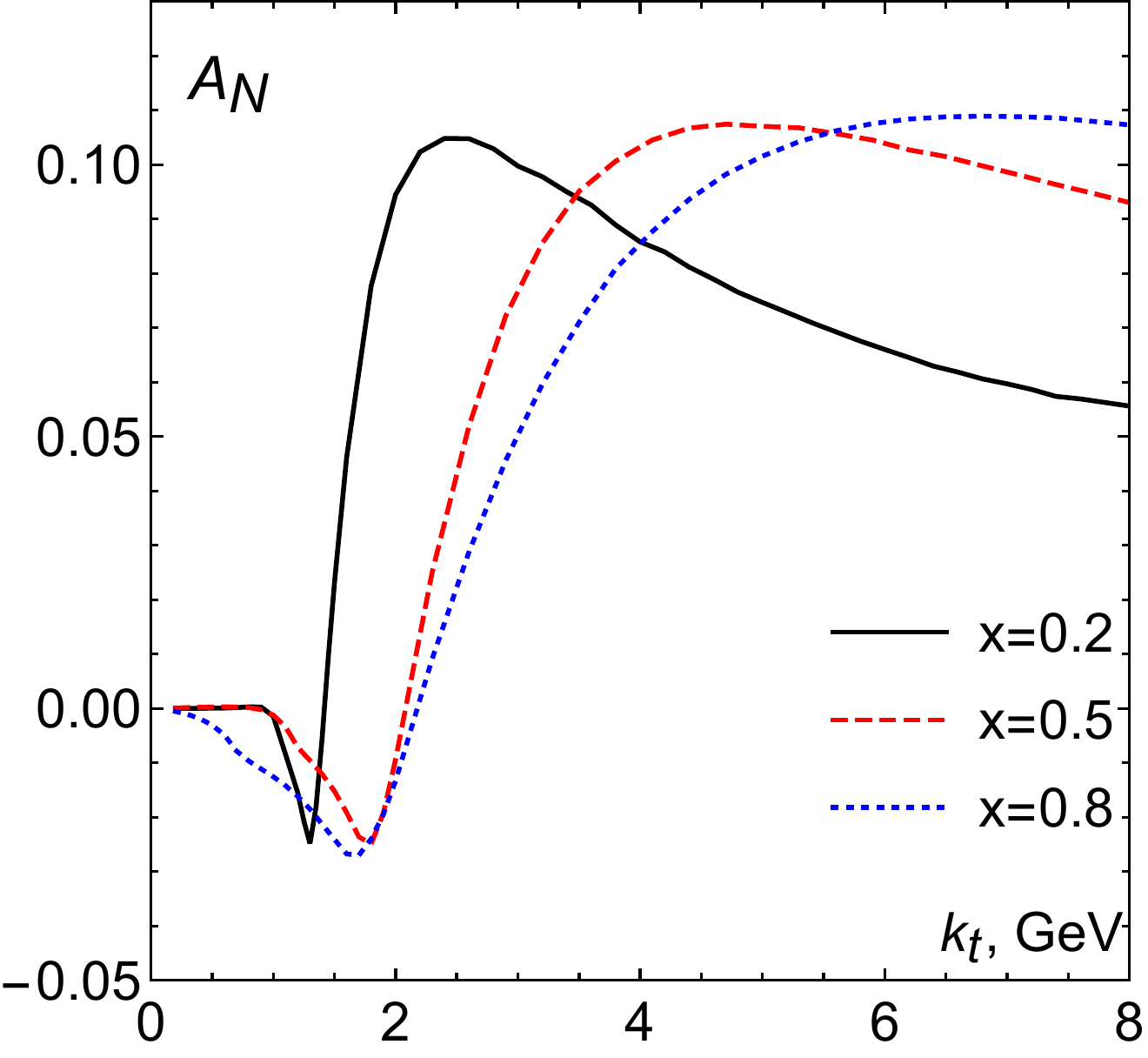}
	\caption{Pion production asymmetry from scattering of constituent quarks. Left: TSSA as function of $x$ for different $x$. Right: TSSA as function of $k_{\perp}$ for $x=0.2,~0.5,~0.8$.}
	\label{fig:ssaQ}
\end{figure}
%
\begin{figure}[tb]
	\centering
	\includegraphics[width=.8\columnwidth]{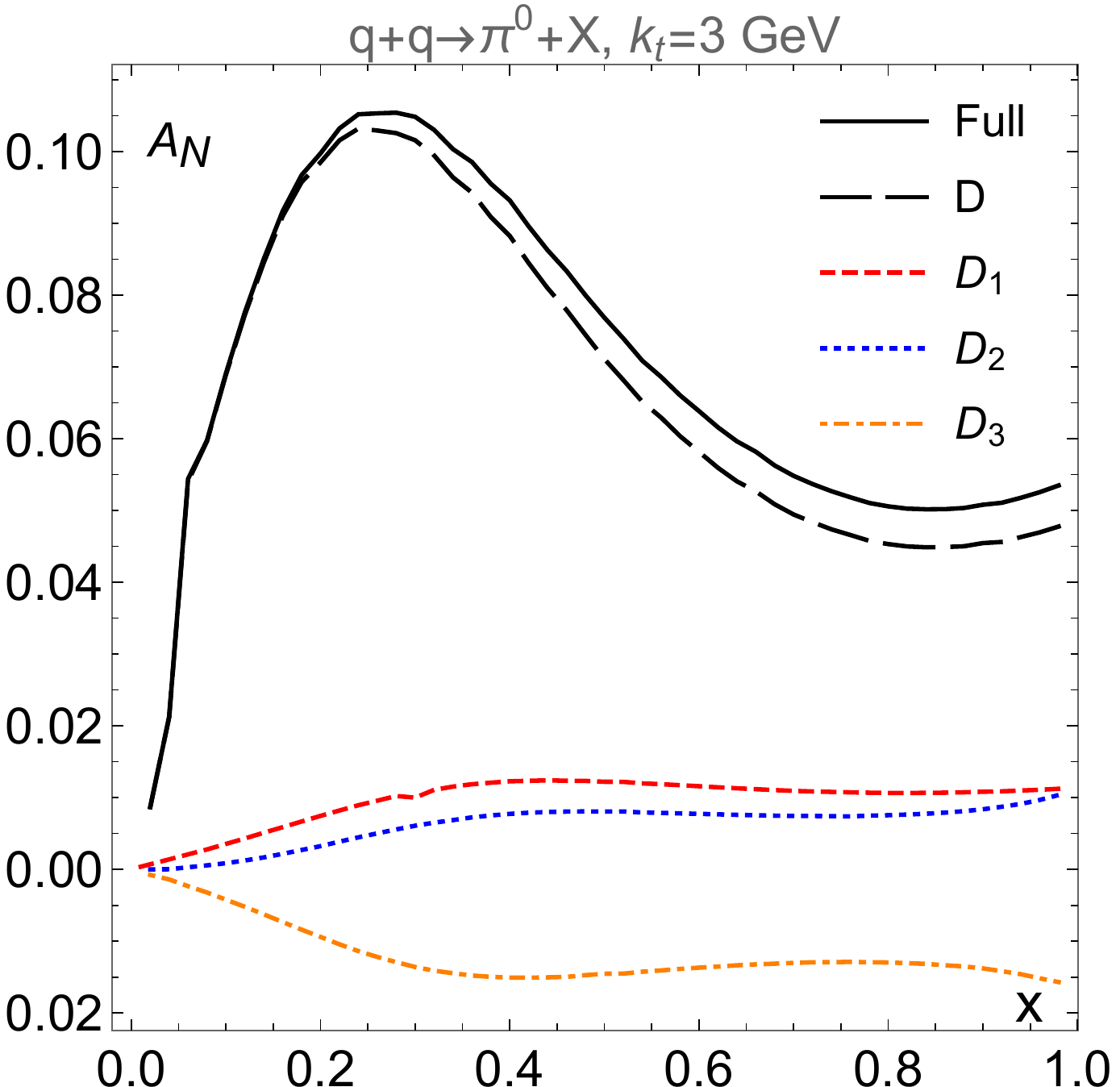}
	\caption{Contribution of diagrams with AQCM vertex on the bottom line to the parton level TSSA. Solid line is total result, long dashed line is result with perturbative bottom vertices. Lines denoted $D_{i}$ correspond to contributions depicted on Fig.\ref{fig:CM_vertex_down_line}.}
	\label{fig:ssaQ_D}
\end{figure}

Fig.\ref{fig:ssaQ} shows asymmetry for parton scattering. 
It is evident that TSSA changes the sign at some $k_t$.
It is due to counteract of two terms in Eq.(\ref{eq:TSSA_parton}): $q_x q_0^2$ and $q_{0x} q^2$.
At small $k_t$ the first term dominates. $q$ grows with $k_t$ and the second term overcomes the first one.
TSSA reach high value $\sim 10\%$ at high $x$ and $k_t$. 
However, for small $k_t$, $A_N$ has peak at smaller $x$ and at bigger $x$ it changes sign.

Fig.\ref{fig:ssaQ_D} demonstrates contribution to TSSA from diagrams with chromomagnetic vertices on the bottom line from Fig.~\ref{fig:CM_vertex_down_line}. This contributions are almost cancel out and the final result does not change significantly. 

In order to calculate hadron cross section and asymmetry, we use set of PDFs provided by NNPDF Collaboration\cite{Nocera:2014gqa}. 
Results on figures are obtained with NLO parton densities(valence + sea quarks) taken at the scale $Q^2=1$ GeV.

\begin{figure}[tb]
	\includegraphics[width=.8\columnwidth]{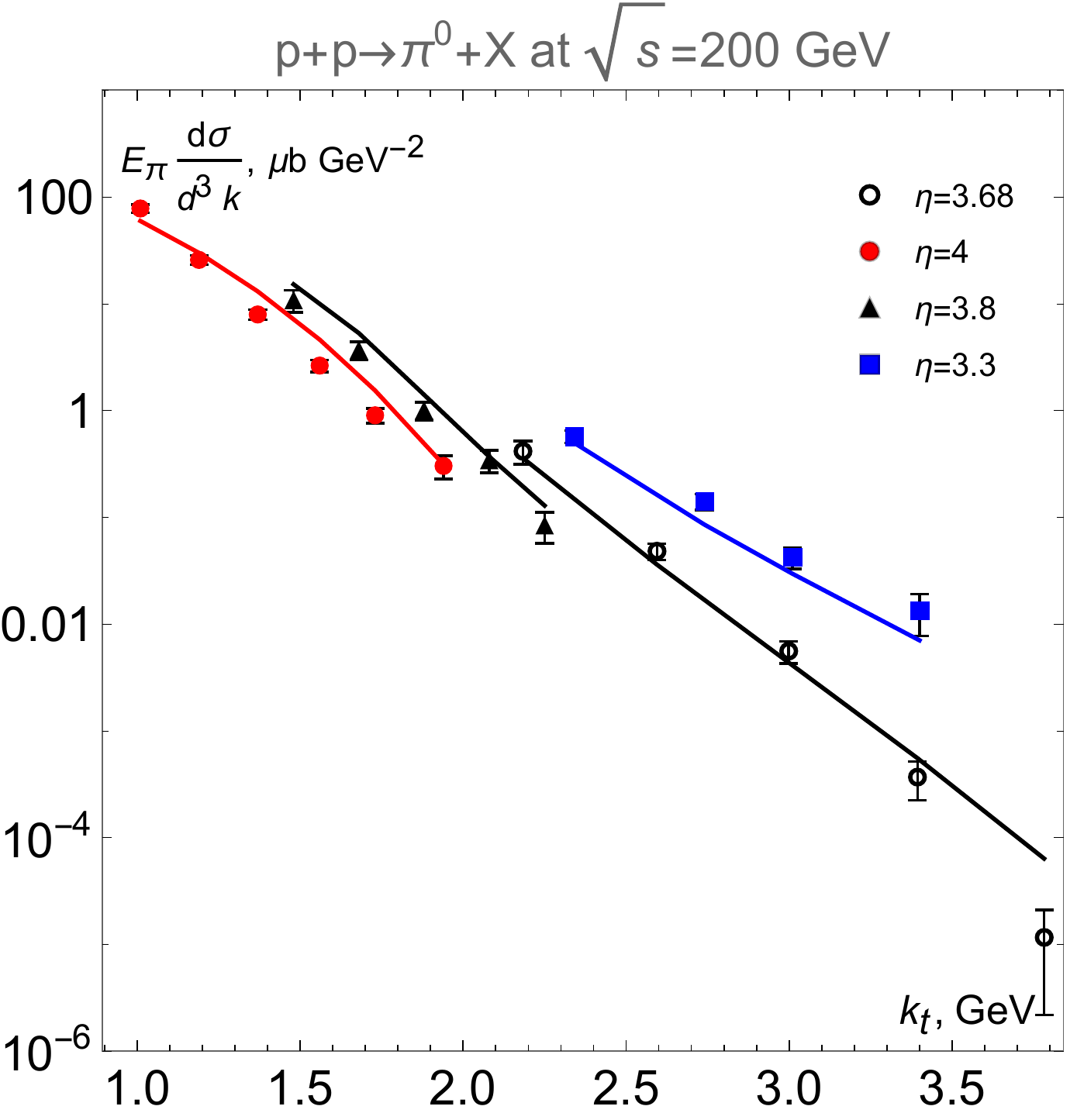}
	\caption{Differential cross section for $\pi^0$ production vs $k_t$ for RHIC. Data are from \cite{Adamczyk:2012xd,Adams:2006uz}}
	\label{fig:cs_rhic}
\end{figure}

Our results for cross section is depicted on Fig.\ref{fig:cs_rhic} and shows agreement with data at RHIC.
Similar pQCD calculations usually are sensitive to a choice of fragmentation functions and scale. 
Good agreement of forward rapidity data and NLO pQCD calculation was reported in \cite{Adamczyk:2012xd}.
However, there DSS fragmentation function\cite{deFlorian:2007aj} has been used, which includes previous RHIC data for fitting.   
Results of calculation with other fragmentation function, which do not include RHIC forward rapidity data to analysis, usually underestimate cross-section by factor 2\cite{deFlorian:2007aj}. 
Overall, for RHIC forward kinematics our model gives predictions similar to pQCD but using less parameters.

Now let's look at TSSA. 
In a non-relativistic framework transverse and longitudinal polarized distributions are equal, $\Delta_T f = \Delta_L f$, since rotations in spin space between different basis commute with spatial operations. 
However, relativistically $\Delta_T f$ and $\Delta_L f$ are different. 
Therefore any difference between helicity and transversity PDFs is related to the relativistic nature of parton dynamics inside hadrons. 
Unfortunately, polarized transverse distribution is poorly known\cite{Radici:2016lam}.
Instead we use the helicity parton densities $\Delta_L f$ from NNPDF as an estimation. 
There are evidences that longitudinal and transverse distributions are the same order\cite{Barone:2001sp,Gockeler:2005cj,Aoki:1996pi}. 
Moreover, nucleon's tensor charge has strong scale dependency and as result the transversity distribution may inherit this strong evolution\cite{Wakamatsu:2008ki,Barone:2001sp}. 
In our estimations we do not consider evolution for transversity and unpolarized pdf.
\begin{figure}[tb]
	\includegraphics[width=0.8\columnwidth]{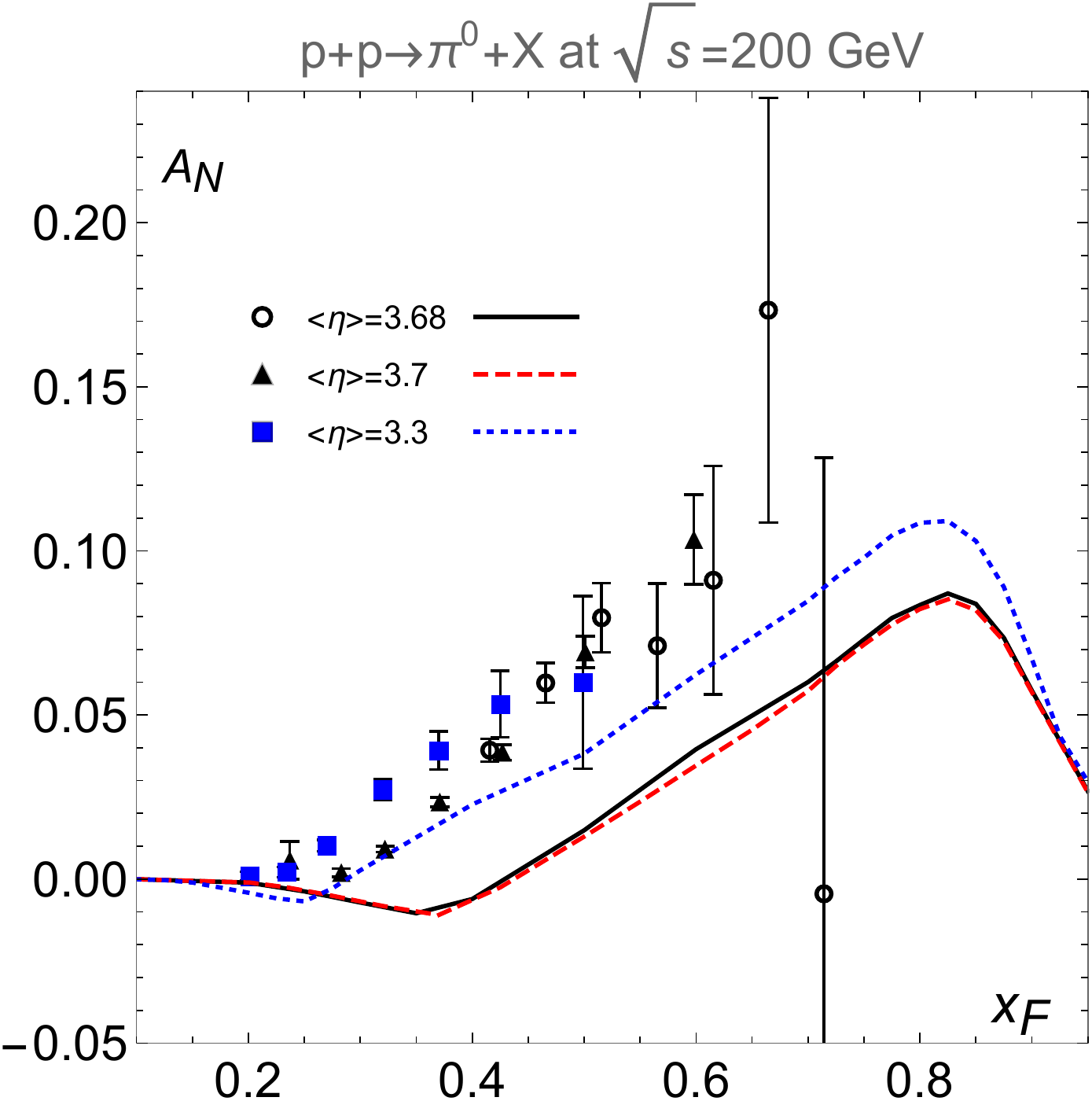}
	\caption{TSSA for $\pi^0$ production on RHIC, data are from \cite{Abelev:2008af,Adamczyk:2012xd}}
	\label{fig:ssa_rhic}
\end{figure}
%
\begin{figure}[tbh]
	\includegraphics[width=1\columnwidth]{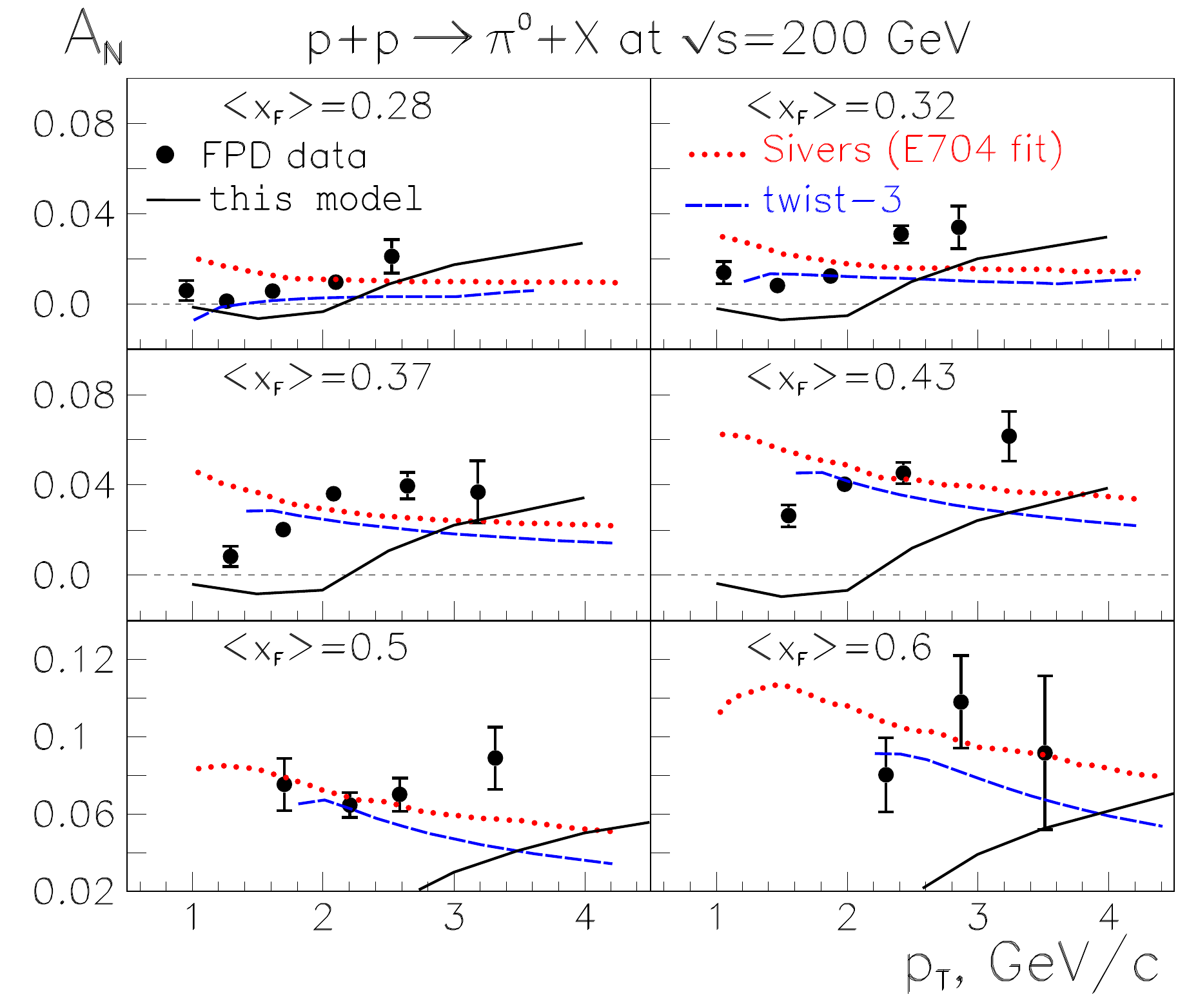}
	\caption{TSSA at individual $x_F$ bins. Data are from \cite{Abelev:2008af}.}
	\label{fig:rhic_ssa_xFbins}
\end{figure}

Fig.\ref{fig:ssa_rhic} shows results for $A_N$ at RHIC energies for the neutral pion. 
Our model predictions are close to data at $\eta=3.3$ and slightly underestimate it.
At higher rapidity discrepancy becomes bigger.
$A_N$ rises with $x_F$ with maximum asymmetry $\approx 10\%$ at $x_F=0.8$. 
Despite that model gives the correct trend of growing asymmetry, theoretical curves are shifted in $k_t$ in comparison with experimental points.
One sees a dependency on pseudorapidity, however data do not have such effect. 
The reason for such behavior in our model is that for the same $x_F$, $k_t$ decreases with $\eta$. 
It is evident from Fig.~\ref{fig:ssaQ} that if $k_t$ is $1-2$~GeV asymmetry becomes small or changes the sign.
This is what happens at $\eta=3.7$ when $x_F\approx 0.4$ on Fig.~\ref{fig:ssa_rhic}. In the case $\eta=3.3$ it occurs at lower $x_F$ and is not so noticeable. 

Fig.\ref{fig:rhic_ssa_xFbins} shows predictions of our model at different $x_F$.
Results from fit of Sivers function \cite{Anselmino:2005ea} and twist-3 fit from \cite{Kouvaris:2006zy} are also shown. 
Notice that our model, in contrast with others, demonstrates asymmetry growing with $k_t$.
Similar to the Fig.\ref{fig:ssa_rhic}, our theoretical curves are shifted to higher $k_t$ in respect with data points. 
A possible reason for ``shifted" results is an interference with other diagrams that we neglected in calculation. This effect requires further study.

An additional contribution to TSSA induced by instantons was
suggested in the papers \cite{Ostrovsky:2004pd} and \cite{Qian:2011ya,Qian:2015wyq}.
It is based on the results from \cite{Moch:1996bs}, where the effects of instantons in the nonpolarized DIS process were calculated.  
In this mechanism the effect arises from phase shift in the quark
propagator in the instanton field. 
This contribution might be complementary to the effect calculated here. Interplay between them could be the reason for overall shift of TSSA to the region of higher $k_t$.

Results for cross-section are sensitive to the value of constituent quark mass $m_q$, because the non-perturbative coupling is proportional to $m_q$. 
In order to describe cross-section data we take $m_q=90$~MeV. 
It is in agreement with Single Instanton Approximation where $m_q=86$~MeV\cite{Faccioli:2001ug}. However, constituent quark masses from the Diakonov-Petrov Model ($m_q=350$ MeV)\cite{Diakonov:2002fq} and Mean Field Approximation($m_q=170$ MeV)\cite{Schafer:1996wv} are too big.

The question how does the proposed mechanism interplay with the factorization approach requires additional study.
In our model fragmentation and hard rescattering are coherent.
It is clear that instanton generated vertices are suppressed at high enough $k_t$, factorization restores and fragmentation must appear from some other process, not coherent with hard rescattering.
If we assume that this incoherent process is completely contained in fitted fragmentation functions, it is impossible to study intermediate kinematic region where both of them at work.
We need a model for fragmentation. 
A possible answer is to calculate fragmentation functions in framework of our model in a way, similar to NJL models\cite{Yang:2016gnd,Nam:2012af}.
If the model gives reasonable results for fragmentation function, it will be possible to study interplay between coherent and incoherent regimes.


\section{Conclusion}\label{section:conclusion}

We calculated TSSA and cross-section for pion production in $pp$ scattering at RHIC energies using the instanton induced effective interaction.
The proposed framework requires less parameters in comparison with the  traditional pQCD approach where one needs parameterize and fit the pion fragmentation function.


Predictions of the model for cross section are consistent with experimental data.
Our model produces the big asymmetry at RHIC kinematics, same magnitude as in experiment. 
However it is shifted to the region of higher $k_t$ in respect to data.
Remarkable outcome of our approach is increase of the asymmetry with transverse momenta of a final particle at given kinematics.
This grow is replaced by a slow decrease at $k_t>5$~GeV.
Such behavior comes from a rather soft power-like form factor of effective vertices and a small average size of instanton, $\rho_c \approx 1/3$~fm, in QCD vacuum.
Similar dependence of asymmetry in $k_t$ is seen in experiment and was not expected in the models based on TMD factorization and ad hoc parametrization of Sivers and Collins functions. 

Another feature of the approach is that $A_N$ does not depend on c.m. energy. The energy independence of TSSA is observed experimentally and in contradiction with naive expectation that spin effects in strong interaction should vanish at high energy.
Moreover, the sign of the TSSA is defined by the sign of AQCM.

Proposed mechanism breaks factorization and can not be treated as an additional contribution to the Sivers distribution function or to the Collins fragmentation function.
In framework of this model, asymmetry in SIDIS and $pp$ is generated by distinct diagrams and in general could be different.  
If this effect has place, Sivers and Collins functions are not universal at small transversal momenta. 
This phenomenon requires further study.

\begin{acknowledgments}
In memory of N.I. Kochelev, a great mentor and scientist, who proposed an idea of this work.
The study was supported by the National Natural Science Foundation of China, Grants No. 11975320(P.M.Z.) and No.11875296 (N.K.). 
N.K. thanks the Chinese Academy of Sciences President’s International Fellowship Initiative for the support via Grants No. 2020PM0073.
\end{acknowledgments}

%
%
\appendix
\section{Phase space}\label{appendix:PS}

In this appendix we give details of phase space calculation for $3$- and $4$-particle final state.
Although it is standard calculation that can be found in textbooks, in our model phase space is limited.

The phase space for three massless particles with momenta $q_+$, $q_-$ and $k$ is
\begin{equation}
\begin{split}
    d\hat{R}_3 &= \frac{(2\pi)^4}{(2\pi)^{9}} \frac{\delta^4(p_+ + p_- \!\!-k-q_- -q_+) d^3 k d^3 q_+ d^3 q_-}{2 E_{q_+} 2 E_{q_-} 2 E_{k}}
    \\
    &=  \frac{1}{(2\pi)^{5}}  \delta(q_-^2) \frac{d^3 k}{2 E_{k}} \frac{d^3 q_+}{2 E_{q_+}} \\
    &= \frac{1}{(2\pi)^{5}} \delta(k^2) \delta(q_+^2)  \delta(q_-^2)  d^4 k d^4 q_+ ,
\end{split}
\end{equation}
where we used $d^3 p/2E=d^4 p \delta(p^2)$ and delta function to remove integration over $d^4 q_-$.
$E_i$ is the energy of a corresponding particle.

We use the decomposition of momenta vectors on light cone vectors $p_+$ and $p_-$,  Eq.(\ref{eq:momenta_decomposition}).
From the decomposition and energy-momentum conservation we get following relations:
\begin{align}
&k^2 = x \beta_k \hat{s} - k^2_\perp;   \nonumber\\
&(p_- -q)^2 = \alpha (\beta - 1)\hat{s}-q_\perp^2;  \\
&(p_+ +q-k)^2 =(1+\alpha-x)(\beta-\beta_k)\hat{s}-(q_\perp - k_\perp)^2;   \nonumber \\
&\alpha_+ = 1+ \alpha - x \approx 1-x ;  \nonumber \\
&\beta_- = 1 -\beta  \approx 1 ;  \nonumber \\
&\alpha = -\alpha_- ;  \nonumber \\
&\beta = \beta_k + \beta_+ . \nonumber
\end{align}
$\perp$ denotes 2-dimensional Euclidean vectors which are transverse to the beam axis $z$. 
Using this decomposition, we rewrite $d\hat{R}_3$ as
\begin{align}
  d\hat{R}_3 =& \frac{\hat{s}^2}{4(2\pi)^5} \frac{dx \, d^2\!k_{\perp} \, d^2\!q_{\perp}}{|x \alpha_+ (\beta - 1) \hat{s}^3|}
  \approx 
  \frac{1}{4(2\pi)^5} \frac{dx \, d^2\!k_{\perp} \, d^2\!q_{\perp}}{x (1-x) \hat{s}}, 
\end{align}
where $d\beta \approx d\beta_+$ has been used.

The next step is to change the integration variable $d^2q_\perp \to d M_{k}^2 d\phi$. $M_k$ is the invariant mass of the pion $k$ and $q_+$ quark:
\begin{align}
  M_k^2 & = (p_+ + q)^2 = (1+\alpha)\beta \hat{s} - q_\perp^2.
\end{align}
Using
\begin{align}
  \alpha \ll 1; \qquad \beta = \beta_+ + \beta_k & = \frac{q_{+\perp}^2}{(1-x+\alpha)\hat{s}} + \frac{k_\perp^2}{x \hat{s}}
\end{align}
we get that
\begin{align}
  M_k^2 & = \frac{(q_{\perp}-k_\perp)^2}{(1-x)} + \frac{k_\perp^2}{x } - q_\perp^2 = \frac{(x q_\perp - k_\perp)^2}{x(1-x)}.
\end{align}

If we define a new perpendicular vector $\tilde{q}_\perp$ as
\begin{equation}
\tilde{q}_\perp^2=(x q_\perp - k_\perp)^2,
\end{equation}
we easily can change the integration variable:
\begin{align}
    d^2q_\perp &= \frac{d^2 \tilde{q}_\perp}{x^2} = \frac{(1-x)}{x} \int_0^{E^2_{\text{sph}}} \!\!\!\!\!\!dM_k^2 \int_0^\pi \!\!\!\!\! d\tilde{\phi}.
\end{align}
$E_{\rm sph}=\frac{3\pi}{2\rho_c}$ is the sphaleron energy which determines the height of potential barriers between different vacuums.
The sphaleron energy restricts allowed phase space.
The final result for the 3-particle phase space is
\begin{equation}
  d\hat{R}_3= \frac{1}{2^7 \pi^5} \frac{dx \, d^2\!k_{\perp}}{x^2 \hat{s}}\int_0^{E^2_{\text{sph}}} \!\!\!\!\!\!dM_k^2 \int_0^\pi \!\!\!\!\! d\tilde{\phi}.
\end{equation}

Next, we need the 4-particle phase space for diagram Fig\ref{fig:Xsection_contributions}(c). It is given by
\begin{align}
    d\hat{R}_4 &= \frac{(2\pi)^4}{(2\pi)^{12}} \frac{\delta^4(p_+ + p_- -l-k-q_- -q_+) d^3 k d^3 l d^3 q_+ d^3 q_-}{2 E_{q_+} 2 E_{q_-} 2 E_{k} 2 E_{l}}.
\end{align}
Using following relations
\begin{align}
  &l^2 = z \alpha_l \hat{s} - l^2_\perp ; & &k^2 = x \beta_k \hat{s} - k^2_\perp;  \nonumber\\
  &\alpha_+ = 1+ \alpha - x \approx 1-x ; && \beta_- = 1 -\beta - z \approx 1-z ;  \nonumber \\
  &\alpha = -\alpha_- - \alpha_l ;& & \beta = \beta_k + \beta_+ ; \\
  &\beta_+ = \frac{q_{+\perp}^2}{\alpha_+ \hat{s}}; &&  q_{+\perp} = q_\perp - k_\perp ;  \nonumber\\
  & \beta_k = \frac{k^2}{x \hat{s}}, \nonumber
\end{align}
we rewrite the expression for the phase space as
\begin{equation}
\begin{split}
  d\hat{R}_4 &= \frac{\hat{s}^3}{2^3 (2\pi)^8}  \frac{dx \, d^2\!k_{\perp} \,  dz\, d^2\!l_{\perp}\,  d^2\!q_{\perp}}{\hat{s}^4 |\alpha_+ \beta_- z x|} \\ 
  &\approx \frac{dx dz d^2 \! k_\perp \, d^2 \! l_\perp \, d^2\! q_\perp}{8 (2\pi)^8 \hat{s} x (1-x) z (1-z)}.
\end{split}
\end{equation}

Now we are going to change the integration variable $d^2l_\perp \to d M_{l}^2 d\phi$, where $M_{l}$ is the invariant mass of the $l$ pion and $q_-$ quark system. Notice that here we replace $d^2l_\perp$, not $d^2q_\perp$.
\begin{equation}\label{Mx_def}
\begin{split}
  M_{l}^2&=(p_- - q)^2=-\alpha(1-\beta)\hat{s}-q_\perp^2 \approx (\alpha_- + \alpha_l)\hat{s}-q_\perp^2 \\
  &=\frac{(z q_{-\perp} - (1-z)l_\perp)^2}{z (1-z)}-q_\perp^2=\frac{(z q_\perp + l_\perp)^2}{z(1-z)},
\end{split}
\end{equation}
where we used $\alpha_- = \frac{q_{-\perp}^2}{\beta_- \hat{s}}\approx\frac{(q_\perp+l_\perp)^2}{(1-z)\hat{s}}$ and $\alpha_l=\frac{l_\perp^2}{z\hat{s}}$. 
If we define a new momentum $n_\perp = z q_\perp + l_\perp$, then we can change integration variable:
\begin{equation}
  \int \!\!\! d^2\! l_\perp = \int \!\!\! d^2 \! n_{\perp} = \int^{2\pi}_0 \!\!\!\!\! d\phi_n \int \frac{dn_{\perp}^2}{2} =  z(1-z) \pi \int\limits_0\limits^{E^2_{\text{Sph}}} \!\! dM^2_{l},
\end{equation}
where integration over the angle has been performed, because the amplitude does not depend on it. The final result is
\begin{equation}
  d\hat{R}_4 = \frac{dx dz d^2 \! k_\perp \, d^2\! q_\perp \, dM^2_{l}}{2^{11} \pi^7 \hat{s} x (1-x)}.
\end{equation}
Later, in analogy with $d\hat{R}_3$ case, we can replace $d^2q_{\perp} \to dM_{k}^2d\tilde{\phi}$.

\section{Amplitudes and parton cross sections}\label{appendix:amplitudes}

In this appendix we give the details of amplitudes and cross section calculation. The expression for the amplitude  shown on Fig.\ref{fig:Xsection_contributions}(a) is
\begin{align}
   |\mathcal{M}_{(a)}|^2 &=  \sum_{f}\overline{\sum_{s,c}} g_s \frac{C(q^2)}{F_\pi} 
   (\bar{u}_{q_+} t^a \sigma_{\mu \nu} q_\nu  \gamma_5 u_{p_+}) \nonumber \\
   &\times (\bar{u}_{q_-} i \gamma_\nu t^{a'} u_{p_-}) 
   D_{\mu\nu}^{aa'}(q) \times \Big[ h.c. \Big],
   \\
   C(q^2) &= g_s \frac{\mu_a}{2 m_q} F_g(q^2)
   = - \frac{3 \pi^{3/2} \rho_c^2 m_q}{4 \sqrt{\alpha_s(\rho_c)}} F_g(q^2), 
   \\
   D_{\mu\nu}^{aa'} &= -i\frac{g_{\mu\nu}\delta^{aa'}}{q^2}.
\end{align}
Note that $\alpha_s$ in $C(q^2)$ is taken at the instanton size scale.
$g_s$ in perturbative vertex is taken at scale $q^2$. We omit writing $q^2$ dependency further.

For a forward scattering, for simplicity of calculation, we use Gribov's decomposition for $g_{\mu\nu}$:
\begin{equation}
g_{\mu\nu}=\frac{2 p_{+\mu} p_{-\nu}}{\hat{s}}+\frac{2 p_{+\nu} p_{-\mu}}{\hat{s}}+g^{\perp}_{\mu\nu} \approx \frac{2 p_{+\mu} p_{-\nu}}{\hat{s}}.
\end{equation}
It allows us to make the following replacement:
\begin{equation}
  D_{\mu\nu}^{aa'} = -i \frac{2 p_{+\mu} p_{-\nu}}{\hat{s}} \frac{\delta^{aa'}}{q^2}.
\end{equation}
That replacement isolates the leading contributions to the amplitude in power of $\hat{s}$.
It leads to the substitution in trace formulas:
\begin{equation}
\begin{split}
\gamma_\mu \rightarrow \cancel{p}_-; &\quad \sigma_{\mu\nu}q_\nu \rightarrow \cancel{p}_-\cancel{q}_\perp \quad \text{for the upper fermionic line,} \\
\gamma_\mu \rightarrow \cancel{p}_+; &\quad \sigma_{\mu\nu}q_\nu \rightarrow \cancel{p}_+\cancel{q}_\perp \quad \text{for the bottom line},
\end{split}
\end{equation}
which factorize traces over fermion lines.
Using it we get
\begin{align}
|\mathcal{M}_{(a)}|^2&= \sum_{f} \frac{1}{4}\frac{\text{Tr}[t^a t^b] \text{Tr}[t^{a'} t^{b'}] \delta^{aa'}\delta^{bb'}}{9} g_s^2 \frac{C^2}{F_{\pi}^2} \frac{4}{\hat{s}^2} \frac{1}{q^4}  \nonumber \\
&\times
  \text{Tr}\Big[ \cancel{q}_+ (\cancel{p}_- \cancel{q} \gamma_5) \cancel{p}_+ (- \gamma_5 \cancel{q} \cancel{p}_-) \Big] 
  \text{Tr}\Big[ \cancel{q}_- \cancel{p}_+  \cancel{p}_- \cancel{p}_+ \Big] \nonumber \\
   &= \sum_{f} \frac{2}{4 \cdot 9} g_s^2 \frac{C^2}{F_{\pi}^2} \frac{16 \hat{s}^2 (1-x)q_{\perp}^2}{q_{\perp}^4},
\end{align}
where the traces are
\begin{align}
  &\text{Tr}\Big[ \cancel{q}_+ (\cancel{p}_- \cancel{q}_{\perp} \gamma_5) \cancel{p}_+ (- \gamma_5 \cancel{q}_{\perp} \cancel{p}_-) \Big] =
   q_{\perp}^2 \text{Tr} \Big[ \cancel{q}_+ \cancel{p}_- \cancel{p}_+ \cancel{p}_- \Big] \nonumber
   \\
  &= 2 q_{\perp}^2 (1-x+\alpha)\hat{s}^2 \approx 2 q_\perp^2 (1-x)\hat{s}^2, \\
  &\text{Tr}\Big[ \cancel{q}_- \cancel{p}_+  \cancel{p}_- \cancel{p}_+ \Big] 
  = 4\cdot 2 q_- p_+ \frac{\hat{s}}{2}
  =2(1-\beta)\hat{s}^2 \approx 2 \hat{s}^2.
\end{align}
We keep the sum over flavor to indicate, that expressions for $\pi^\pm$, $\pi^0$ are different.

In case with nonperturbative vertex at bottom line Fig.\ref{fig:Xsection_contributions}(b), the corresponding trace is:
\begin{equation}
\begin{split}
  \text{Tr}\Big[ \cancel{q}_- \cancel{p}_+ \cancel{q}_{\perp}  \cancel{p}_- \cancel{q}_{\perp} \cancel{p}_+ \Big] &=
q_{\perp}^2 \text{Tr}\Big[\cancel{q}_- \cancel{p}_+  \cancel{p}_-  \cancel{p}_+ \Big]
\\
& =2(1-\beta)\hat{s}^2 q_{\perp}^2 
\approx 2 q_{\perp}^2 \hat{s}^2.
\end{split}
\end{equation}
Repeating similar calculations we get
\begin{align}
|\mathcal{M}_{(b)}|^2& = \sum_{f} \frac{2}{4 \cdot 9} \frac{C^4}{F_{\pi}^2} \frac{16 \hat{s}^2 (1-x)q_{\perp}^4}{q_{\perp}^4}.
\end{align}

The amplitude for the two pion contribution $2q \to 2\pi 2q$ Fig.\ref{fig:Xsection_contributions}(c) is very similar to the case with one pion vertex. 
The difference is only in the trace over the  bottom fermion line, which becomes similar to the upper line.
\begin{equation}
\begin{split}
  |\mathcal{M}_{(c)}|^2 &= \sum_{f}\overline{\sum_{s,c}}  
   -\frac{C^2}{F_{\pi}^2} 
  (\bar{u}_{q_+} \sigma_{\mu \lambda} q_\lambda  \gamma_5 t^a u_{p_+}) \\
  &\times (\bar{u}_{q_-} \sigma_{\nu\rho} q_{\rho} \gamma_5 t^{a'} u_{p_-}) 
 D_{\mu\nu}^{aa'}(q^2) \times \Big[ h.c. \Big] 
 \\
  &=\sum_{f} \frac{C^4}{F_{\pi}^4} \frac{\text{Tr}[t^a t^b] \text{Tr}[t^{a'} t^{b'}]}{4 \times 9} \delta^{aa'}\delta^{bb'}  \frac{4}{\hat{s}^2} 
  \\
  & \times \text{Tr}\Big[ \cancel{q}_+ (\cancel{p}_- \cancel{q}_\perp \gamma_5) \cancel{p}_+ (- \gamma_5 \cancel{q}_\perp \cancel{p}_-) \Big] 
  \\
  & \times \text{Tr}\Big[ \cancel{q}_- (\cancel{p}_+ \cancel{q}_\perp \gamma_5) \cancel{p}_- (- \gamma_5 \cancel{q}_\perp \cancel{p}_+) \Big]
  \\ 
  &=
  \sum_{f} \frac{8 \hat{s}^2 C^4}{9 F_{\pi}^4} q_\perp^4 \frac{(1-x)(1-z)}{q^4_\perp}.
\end{split}
\end{equation}
Final formulas for the parton cross section are
\begin{equation}
  d\hat{\sigma}_{(a)}= \sum_{f}\!\!\int\limits_{0}\limits^{E^2_{\rm sph}} \!\!\! dM_k^2 \!\! \int\limits_{0}\limits^{\pi} \!\!\! d\tilde{\phi} \, \frac{g_s^2 C^2}{9(2\pi)^5 F_{\pi}^2} \frac{1-x}{q_\perp^2 x^2}dx d^2k_\perp,
\end{equation}
\begin{equation}
  d\hat{\sigma}_{(b)}=\sum_{f} \!\! \int\limits_{0}\limits^{E^2_{\rm sph}} \!\!\! dM_k^2 \!\! \int\limits_{0}\limits^{\pi} \!\!\! d\tilde{\phi} \, \frac{ C^4}{9(2\pi)^5F_{\pi}^2} 
  \frac{1-x}{x^2}dx d^2k_\perp.
\end{equation}
with perturbative and chromomagnetic bottom vertices respectively.

In case of two pions we have
\begin{align}\label{eq:2pi_sigma_parton_xsec_1}
  d\hat{\sigma}_{(c)} &= \frac{|\mathcal{M}_{(c)}|^2}{2\hat{s}}d\hat{R}_4 \nonumber
  \\
  &= \sum_{f} \int
  \frac{C^4}{9 F_{\pi}^4} \frac{(1-z)}{2^{9} \pi^7 x}
 dM^2_{l} dx dz d^2k_\perp d^2 q_\perp \\
   &= \sum_{f} \frac{C^4}{9 F_{\pi}^4} 
   \frac{(1-z)}{2^9 \pi^7 x} E^2_{\text{sph}} dx dz d^2k_\perp
   d^2 q_\perp,
\end{align}
where integration over $dM^2_{l}$ was done and we got $E^2_{\text{sph}}$ in the last line.
Next we will integrate out $dz$, which gives the factor $1/2$:
\begin{equation}\label{eq:2pi_sigma_parton_xsec_2}
  d\hat{\sigma}_{(c)} = \sum_{f} \frac{C^4}{9 F_{\pi}^4} \frac{E^2_{\text{sph}} dx d^2k_\perp   d^2 q_\perp}{2^{10} \pi^7 x}.
\end{equation}
Replacing integration over $d^2q_\perp$ by $dM^2_k$ we get:
\begin{equation}
  d\hat{\sigma}_{(c)} = \sum_{f}  \frac{C^4 E^2_{\text{sph}}}{9 \times 2^{10} \pi^7 F_{\pi}^4} \frac{(1-x)}{x^2} \, dx \, d^2k_\perp \,  d M^2_k \, d\tilde{\phi}.
\end{equation}
In the end let us briefly discuss flavor summation. Pion field in flavor space decomposed as
\begin{equation}
  \vec{\tau}\vec{\phi}=\sqrt{2} (\tau_+ \pi^+ + \tau_- \pi^-) + \tau_0 \pi^0,
\end{equation}
\begin{equation}
  \tau^+=\begin{pmatrix}
           0 & 1 \\
           0 & 0 \\
         \end{pmatrix}, \quad
  \tau^- = \begin{pmatrix}
           0 & 0 \\
           1 & 0 \\
         \end{pmatrix}, \quad
  \tau^0 = \begin{pmatrix}
           1 & 0 \\
           0 & -1 \\
         \end{pmatrix}.
\end{equation}
For $\pi^0$
\begin{align}
  |M_{\pi^0}|^2 \propto &\Big( \bar{\psi}_u (...) \psi_u - \bar{\psi}_d (...) \psi_d \Big) \Big( \bar{\psi}_u (...) \psi_u - \bar{\psi}_d (...) \psi_d \Big)^* \nonumber \\
  = &\Big( \bar{\psi}_u (...) \psi_u \Big)^2+\Big( \bar{\psi}_d (...) \psi_d \Big)^2,
\end{align}
where $(...)$ denotes any expression with Dirac $\gamma$ matrices.
For charged pions it is
\begin{align}
  |M_{\pi^{\pm}}|^2 \propto &\sqrt{2} \Big( \bar{\psi}_{u,d} (...) \psi_{d,u} \Big) \sqrt{2} \Big( \bar{\psi}_{u,d} (...) \psi_{d,u} \Big)^* \nonumber \\
  =&2 \Big( \bar{\psi}_{u,d} (...) \psi_{d,u} \Big)^2.
\end{align}

\section{Integration over parton momenta fraction}\label{appendix:int_limits}

A differential hadron cross-section is a convolution of parton distribution functions(PDF) and the partonic cross-section $d\hat{\sigma}$
\begin{equation}
  d\sigma = \int dx_a dx_b \, f(x_a) f(x_b) d\hat{\sigma}
\end{equation}
The momenta of an exclusive hadron in the parton c.m. frame and hadron c.m. frame are related:
\begin{equation}
\begin{split}
  &\hat{k}_z=x\frac{\sqrt{\hat{s}}}{2}; \qquad  k_z=x_F\frac{\sqrt{s}}{2}  
  \\
  &E_k=\sqrt{k_{\perp}^2+k_{z}^2+m_{\pi}^2}\approx\frac{\sqrt{s}}{2}\sqrt{x_T^2+x_F^2} ; 
  \\ 
  &x_{T,F}=\frac{2}{\sqrt{s}}k_{\perp,z}; \quad x_F=x_a x;
  \\
   &k_z=\sqrt{\frac{x_a}{x_b}} \hat{k}_z; \quad k_{\perp} = \hat{k}_{\perp}.
\end{split}
\end{equation}
The hat denotes a value in the parton c.m. frame. Rewriting the cross section in terms of momenta in the hadron frame we get
\begin{equation}
E_k \frac{d^3 \sigma}{d^3k} = \sum_{f} \int^{x_a^{\max}}_{x_a^{\min}} \!\!\!\!\!\! dx_a \int^{x_b^{\max}}_{x_b^{\min}} \!\!\!\!\!\! dx_b \, f(x_a) f(x_b) \frac{2 E_k}{\sqrt{s}x_a} \frac{d \hat\sigma}{dx d^2 k_{\perp}}.
\end{equation}

We have $2\to 3$ or $2\to 4$ parton subprocess. It means that we can not recklessly use $\hat{s}+\hat{t}+\hat{u} =0$, which is true for $2\to2$ subprocess with massless particles. However, we can combine all particles, except the detected one, into an effective particle and reduce our case to $2\to 2$. Denote the mass of the effective particle as $X^2$.
\begin{align}\label{eq:stu_X}
   & \hat{s}+\hat{t}+\hat{u} = X^2,\\
   & X^2=(p_+ + p_- - k)^2 = (1-x)\hat{s}-\frac{k^2_\perp}{x}.
\end{align}
For the cross section with one pion we combine only $q_-$ and $q_+$ quarks. All formulas are valid for this case also, because we can just put $l=0$. Note that $\hat{t}$ is not a gluon virtuality, $\hat{t} \neq q^2$.

We need express parton variables through hadron level variables:
\begin{align}
 &\hat{s} = x_a x_b s; &&  \hat{t} = x_a t;    &\hat{u}= x_b u;
\end{align}
\begin{align}
  x_1 &= -\frac{u}{s}=\frac{x_T}{2 \tan(\theta_h /2)};  \\
  x_2&=-\frac{t}{s}=\frac{x_T \tan(\theta_h /2)}{2}.
\end{align}
$\theta_h$ is the pion scattering angle in the hadron c.m. frame. 

\begin{figure}[bth]
  \includegraphics[width=0.8 \columnwidth]{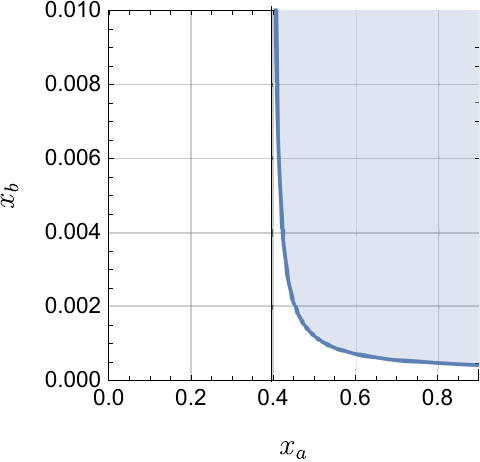}
  \caption{Integration region over $x_a$ and $x_b$(shaded area) for kinematics  $\sqrt{s}=200$~GeV, $k_{\perp}=2$~GeV, $\eta=3.68$. }
  \label{fig:xavsxb}
\end{figure}

To determine maximum and minimum values for $x_a$ and $x_b$, we notice from Eq. \ref{eq:stu_X}
\begin{equation}
x_b= \frac{X^2/s + x_a x_2}{x_a-x_1}
\end{equation}
For fixed $X$, it is monotonically decreasing function with $x_a$. 
Therefore $x_a=x_a^{\min}$ when $x_b=1$,
\begin{align}\label{eq:int_lim_from_stu}
\begin{split}
	x_a^{\min} = \frac{x_1}{1-x_2};          \qquad&x_a^{\max}=1;\\
	x_b^{\min} = \frac{x_a x_2}{x_a - x_1};  \qquad&x_b^{\max}=1.
\end{split}
\end{align}
This limits allow kinematic region where invariant mass $X^2$ becomes negative.
We need additional constrain coming from $X^2>0$:
\begin{equation}\label{eq:Xpositivity}
  x_b> \frac{k^2_{\perp}/x}{x_a(1-x)s}.
\end{equation}
However, it is not important for RHIC kinematics. Integration region is almost identical to one determined by Eq.\ref{eq:int_lim_from_stu} alone.
The Fig.\ref{fig:xavsxb} shows the example of integration region over $x_a$ and $x_b$. 

One may notice that for the amplitude with two pions Fig.\ref{fig:Xsection_contributions}(c) $X^2=(q_{+} + q_{-})^2$, without inclusion of the pion $l$.
As result, limits for $x_{a,b}$ depend on $l_{\perp}$ and $z$. Therefore we can not integrate over $z$ and $l_{\perp}$ independently as we did in (\ref{eq:2pi_sigma_parton_xsec_2}).
That more rigorous calculation was done and the correction is less than $5\%$ for cross section at the RHIC kinematics.
Therefore we can use this approach as a good approximation.


\end{document}